\documentclass[twocolumn]{aastex6}

\usepackage{graphicx}
\usepackage{natbib}
\usepackage{wrapfig}
\usepackage{amsmath}
\usepackage{gensymb}
\usepackage{url}
\usepackage{hyperref}

\usepackage{float}
\usepackage{color}
\usepackage{amssymb}
\usepackage{epstopdf}

\graphicspath{./}
\newcommand{\Msun}{\rm{M_{\odot}}}
\newcommand{{\Gizmo}}{{\small\sc Gizmo}}
\newcommand{\Gadget}{{\small\sc Gadget}}
\newcommand{\Arepo}{{\small\sc Arepo}}

\begin{document}

\title{The formation and evolution of star clusters in interacting galaxies}
	
\author{
Moupiya Maji$^{1, 2}$, Qirong Zhu$^{1,2}$, Yuexing Li$^{1, 2}$, Jane Charlton$^{1}$, Lars Hernquist$^3$ and Alexander Knebe$^{4,5}$  
}

\affil{$^1$Department of Astronomy \& Astrophysics, The Pennsylvania State University, 
525 Davey Lab, University Park, PA 16802, USA\\
$^2$Institute for Gravitation and the Cosmos, The Pennsylvania State University, University Park, PA 16802, USA\\
$^3$ Harvard-Smithsonian Center for Astrophysics, Harvard University, 60 Garden Street, Cambridge, MA, 02138, USA\\
$^4$ Departamento de F\'{i}sica Te\'{o}rica, M\'{o}dulo 8, Facultad de Ciencias, Universidad Aut\'{o}noma de Madrid, 
E-28049 Madrid, Spain\\ 
$^5$ Astro-UAM, UAM, Unidad Asociada CSIC, E-28006 Madrid, Spain}

\email{Email:moupiya@psu.edu}

\begin{abstract}

Observations of globular clusters show that they have universal
lognormal mass functions with a characteristic peak at $\sim 2\times
10^{5}\, \Msun$, but the origin of this peaked distribution is highly
debated. Here we investigate the formation and evolution of star
clusters in interacting galaxies using high-resolution hydrodynamical
simulations performed with two different codes in order to mitigate
numerical artifacts.  We find that massive star clusters in the range
of $\sim 10^{5.5} - 10^{7.5}\, \Msun$ form preferentially in the
highly-shocked regions produced by galaxy interactions. The nascent
cluster-forming clouds have high gas pressures in the range of $P/k
\sim 10^8 - 10^{12}\, \rm{K cm^{-3}}$, which is $\sim 10^4 - 10^8$ 
times higher than the typical pressure of the interstellar medium but consistent with recent observations
of a pre-super star cluster cloud in the Antennae Galaxies.
Furthermore, these massive star clusters have quasi-lognormal initial
mass functions with a peak around $\sim 10^{6}\, \Msun$. The number of
clusters declines with time due to destructive processes, but the
shape and the peak of the mass functions do not change significantly
during the course of galaxy collisions. Our results suggest that
gas-rich galaxy mergers may provide a favorable environment for the
formation of massive star clusters such as globular clusters, and that
the lognormal mass functions and the unique peak may originate from
the extreme high-pressure conditions of the birth clouds and may
survive the dynamical evolution.

\end{abstract}


\section{Introduction}
\label{sec:intro}

Star clusters (SCs) are building blocks of galaxies, so their origin and
evolution are important aspects of the study of galaxy formation.
Over the past two decades, numerous young massive clusters (YMCs)
have been observed by the Hubble Space Telescope in interacting and
merging galaxies, such as NGC 1275 \citep{Holtzman1992}, NGC 7252
\citep{Whitmore1993}, NGC 3921 \citep{Schweizer1996}, NGC 4038/39 or
the Antennae Galaxies \citep{Whitmore1995, Whitmore1999,
  Whitmore2010}, NGC 4449 \citep{Annibali2011}, and NGC 7176/7174
\citep{Miah2015}. The YMCs formed in these environments are compact
($\sim$ few parsecs), gravitationally-bound objects with masses $>10^4\,
\Msun$ and ages $\sim 10 - 100$ Myr \citep{Zwart2010}.  The initial cluster
mass function (ICMF) of these young clusters, however, is not
well-determined. Some studies suggested that it can be described as a
falling power law with $dN/dM \propto M^{-2}$ \citep{Zhang1999,
  BiK2003, McCrady2007, Fall2012}, some argued that it might be better
fit by a Schechter function with power index -2 and a characteristic
mass of few $10^6\, \Msun$ \citep{Bastian2008, Zwart2010}, and some
proposed that it is not a power law at all mass scales but has a
turnover at the low mass end \citep{Cresci2005, Anders2007}.

On the other end of the SC spectrum are old globular
clusters (GCs) that have been observed extensively in nearby galaxies
\citep[e.g.,][]{Harris1991, Brodie2006, Gratton2012, Kruijssen2014,
  Kruijssen2015}. These are massive ($\sim 10^4 - 10^6\, \Msun$),
gravitationally-bound, compact (few pc) and old (age $\simeq 10 -13$
Gyr) systems that formed in the early universe and have survived to
the present-day \citep{Forbes2010, VandenBerg2013}. It has been widely
suggested that YMCs could be progenitors of these globular clusters
\citep{deGrijs2007, Longmore2014}. However, the observed globular
cluster mass functions (GCMFs) are bell-shaped or lognormal-shaped
with a peak mass around $1.5 - 3 \times 10^5\, \Msun$
\citep{Harris2001, Jordan2007}, which are remarkably different from
those of YMCs.

In order to explain the discrepancy between mass functions of YMCs and
GCs, a number of studies have focused on the dynamical evolution of
SCs. Some theorized that the GCs were formed from the
collapse of protogalactic clouds and these clusters had bell shaped
ICMFs to begin with \citep{Fall1985, Vesperini2000, Vesperini2001,
  Parmentier2007}. The more prevalent theory suggested that young GCs
start with a power-law ICMF, and during evolution they are affected by
a number of destructive processes that can disrupt the lower mass
clusters more easily and more frequently than their higher mass
counterparts, resulting in a lognormal profile \citep{Gnedin1997,
  Baumgardt1998, Fall2001}. Such destruction can rise from two-body
relaxation, shock heating, supernova explosions, tidal shocking and stellar dynamical
evaporation \citep[e.g.,][]{Gnedin1999a, Fall2001, McLaughlin2008}. In particular,
tidal forces induced by galaxy interactions or GCs passing through a
galactic disk can generate efficient heating from strong tidal shocks, which
significantly affect the evolution of GCs \citep{Combes1999,
 Gnedin1999a, Gnedin1999b}.

However, little is known about the formation conditions that determine
the mass functions of YMCs and how they are related to those of
GCs. It has long been suggested that globular clusters preferentially
form in regions with extremely high pressure \citep{Elmegreen1997,
  Ashman2001}. High pressure in molecular clouds can result in high
velocity dispersions (several tens of km/s) which lead to larger
binding energy. This helps the cloud not to get dispersed by typical
HI clouds. With rising pressure, the specific star formation
efficiency of the region can increase significantly by up to one order
of magnitude \citep{Jog1992, Jog1996}. High binding energy and high
specific star formation efficiency are critical to the formation of
massive, bound SCs. From the present-day properties of GCs,
it is suggested that the cluster-forming clouds should have
experienced high pressure on the order of $P/k_B \gtrsim 10^8\, \rm{K}
cm^{-3}$, which is $\gtrsim 10^4$ times larger than the ambient
interstellar medium pressure in our galaxy \citep{Jenkins1983, Boulares1990, Elmegreen1997, Welty2016}.
However, these extreme pressures can be easily produced in interacting
galaxies by violent shocks, so they are theoretically expected to be
ideal formation sites for GCs.

On the observational front, it has been difficult to directly observe
the physical conditions of a proto super-star cluster cloud (SSC - star clusters with the possibility of evolving
into GCs). \cite{Wei2012} observed molecular cloud regions in the Antennae
Galaxies and found very massive ($\gtrsim 10^6\, \Msun$) clouds in the
centers of high star formation regions with large velocity
dispersion. Recently, \cite{Johnson2015} have studied the properties
of a pre-SSC cloud in the merging galaxies of
the Antennae in detail via CO observations. This cloud is not yet
forming stars, but is expected to begin doing so in less than 1 Myr,
which makes it an ideal candidate to investigate SSC formation
conditions.  Direct measurements of the cloud suggest that it has mass
of $ > 5\times 10^6\, \Msun$ and a radius of $\sim 25$ pc which falls in
the range of GC properties. The cloud is experiencing a tremendously
high external pressure $P/k_B > 10^8$ K cm$^{-3}$ . \cite{Adamo2015}
studied the SCs in M83 at different radii from the galaxy
center and concluded that high gas pressure increases cluster
formation efficiency.

In order to investigate the formation and evolution of SCs
and their mass functions, we need realistic simulations of galaxies
with SC systems to understand the complex interplay of all
the creation and destruction processes. However, due to the large dynamical range (from sub
pc for star formation to kpcs for galaxies), mass scale (from star
clusters of $\sim 10^4\, \Msun$ to galaxies of $10^{12}\, \Msun$), and
the various physical processes involved (cluster formation in GMCs,
stellar evolution, binary interaction, shocks, tidal disruptions etc),
it has been a challenge to study formation and evolution of SCs in galaxies
numerically. Most of the early simulation efforts assumed a shape for
the ICMF, generally power laws or Schechter functions, and
then simulated their evolution using N-body codes
\citep{Vesperini1997, Baumgardt2003, Lamers2010}. Some simulations
have focused on particular aspects of the problem, such as the
evolution of GCs in the tidal fields of mergers \citep{Renaud2013},
star escape rate from GCs \citep{Gieles2008} and the effects of
intermediate mass black holes on GCs \citep{Lutzgendorf2013}. A few
simulations explore SCs in specific environments such as
high redshift galaxies \citep{Prieto2008} and dwarf galaxies
\citep{Kruijssen2012a}. Some variants of N-body simulations have
also been applied, for example, \cite{Renaud2011} used a tensor field
to describe tidal fields.

It is very important to include hydrodynamics of the gas in the galaxy
to fully understand SC formation and evolution, but it can
be highly computationally expensive to explore the entire range of
processes.  For example, \cite{Li2004} used sink particles to
represent SCs in high-resolution, smoothed particle
hydrodynamics (SPH) simulations but could not follow the structure of
clusters; \cite{Kruijssen2011, Kruijssen2012b} used N-body/hydro
simulations for the galaxies but followed the cluster evolution
semi-analytically.  A more complete treatment of galaxy simulation 
and SC identification emerged recently. \cite{Renaud2015}
modeled an Antennae-like merger using an adaptive mesh refinement (AMR)
grid-based code and identified SCs with a friends-of-friends
(FOF) group finding algorithm. They found that the cluster formation
rate roughly follows the star formation rate, and that clusters formed
in interacting galaxies are up to 30 times more massive than those
formed in isolated galaxies. However, a detailed study of the
formation conditions of SCs and the evolution of cluster
mass functions is needed.

In this study, we perform fully hydrodynamic simulations of galaxy
mergers using two different codes: {\Gadget} \citep{Springel2001, Springel2005} and {\Gizmo} \citep{Hopkins2015}. 
We identify
the SCs in them as overdense groups of bound particles,
using the Amiga Halo Finder (AHF\footnote{The AHF code is available at \url{http://popia.ft.uam.es/AHF/Download.html}}, 
\citealt{Gill2004, Knollmann2009}). We investigate the physical conditions of
SC formation by tracking the properties of the nascent birth
clouds. We follow the early evolution of their mass function to
understand the connection between YMCs and GCs and the origin of the
mass function peak of GCs. This is one of the first studies to
realistically identify SCs and follow their formation and
evolution in galaxy mergers.

Our paper is organized as follows: in \S~\ref{sec:methods} we describe
the methods, which include the numerical codes, galaxy model and
cluster identification; in \S~\ref{sec:formation} we present the
results of cluster formation and physical conditions, and the initial cluster mass functions; in
\S~\ref{sec:evolution} we explore the evolution of cluster mass
functions; in \S~\ref{sec:discussions} we discuss the limitations of
our study; and we summarize our findings in \S~\ref{sec:conclusions}.

\section{Method}

\label{sec:methods}

In this study, we perform hydrodynamical simulations of a galaxy
merger of two Milky Way-size progenitors using two different hydrodynamics
codes: the improved SPH code {\Gadget} developed by
\cite{Springel2001} and \cite{Springel2005}, and the new meshless code
{\Gizmo} developed by \cite{Hopkins2015}. In order to reduce
numerical artifacts on the physical results, we have implemented
the same physical processes in both codes, and use the same
initial conditions in the simulations. The SCs are
identified in the simulations using a density-based group
finding algorithm Amiga Halo Finder (AHF, \citealt{Knollmann2009,
Gill2004}). In what follows we briefly describe the codes,
galaxy model and SC identification used in the
simulations; we refer the reader to read the references therein
for detailed descriptions.

\subsection{Hydrodynamic Codes}
\label{subsec:codes}

{\Gadget} \citep{Springel2001, Springel2005} is a massively parallel
N-body/SPH code. It handles the components of a galaxy in two distinct
ways: it treats the motions and evolution of dark matter and stars as
collisionless particles in an N-body problem, while the gas is dealt
with using the SPH method \citep{Gingold1977, Hernquist1989}. The
N-body particles are described by the collisionless Boltzman and
Poisson equations, and the hydrodynamics of the fluid is followed
using properties of neighboring gas particles smoothed by a kernel
function. The gravitational force of each particle is calculated with
a tree algorithm in which particles are grouped together and their
effect is taken as a single multipole force, which reduces the
computation cost greatly to O($N log N$) compared to the direct
summation of each particle pair with complexity O($N^2$). In this code, an artificial 
viscosity
term is introduced into the equation of motion of SPH to represent the
viscosity which often arises in ideal gases due to shocks caused by
microphysics. The {\Gadget}-2 we use explicitly conserves energy and
entropy in the SPH formulation \citep{Springel2002}.  This version and
its variants have been widely used in a large number of applications,
from large-scale cosmological simulations
\citep[e.g.,][]{Springel2003a, Springel2005millenium, Feng2013,
  Schaye2015} to galaxy mergers \citep[e.g.,][]{Springel2000, Li2004,
  Springel2005merger, Hopkins2005, Hopkins2006, Hopkins2008,
Li2007, Cox2008, Hayward2014}.

{\Gizmo} \citep{Hopkins2015} is a new Lagrangian code developed to
circumvent the many problems encountered by SPH methods \citep{Agertz2007, 
Bauer2012, Vogelsberger2012, Sijacki2012, Keres2012, Nelson2013, Zhu2015, Zhu2016}. 
It derives
the hydrodynamic equations using a kernel function to partition the
volume, and a Riemann solver to evolve the equations at the Lagrangian
face co-moving with the mass. {\Gizmo} implements strict conservation
of mass, energy and linear and angular momentum and it does not
require any artificial diffusion terms to deal with shocks, embodying
the advantages of both SPH and grid-based methods.  It captures the
instabilities of fluid mixing well, greatly reduces numerical noise
and artificial viscosity and as a result calculates fluid physics at
smaller Mach number more accurately. {\Gizmo} treats the contact
discontinuities and shocks more precisely and more efficiently,
generally within one kernel length instead of 2-3 as in {\Gadget}, and
it does not have the zeroth order and first order errors that are
present in SPH \citep{Zhu2015}, so it can attain higher accuracy with
a much smaller number of neighbors which results in a faster
convergence. We have used the meshless finite-mass mode of Gizmo for our project.
The mass of an individual gas element is conserved in this mode, 
which allows us to trivially trace
the history of star particles
to their progenitor gas particles 
(otherwise, one needs tracer particles to do so).

A detailed comparison between {\Gadget} and {\Gizmo} in galaxy
simulations has been conducted by \cite{Zhu2016}, who showed a general
agreement between the two simulations but there were notable
differences in a number of galaxy properties such as star formation
history, gas fraction and disk structures.

In this study, our motivation for using these two codes to perform the
same merger simulation is to reduce the possibility of numerical artifacts
affecting our results. As we will show in
\S~\ref{sec:formation}, although the detailed star formation history 
of the mergers is different between the two simulations, the overall
distribution functions of the cluster mass and the pre-cluster gas
pressure agree well, which suggests that our results are physical and
robust, because we can argue against the effects of numerical artifacts
such as the number of neighbors and artificial viscosity on these
results.

\subsection{Galaxy Model}

Our simulations consider major mergers of two equal-mass Milky Way - sized
galaxies. The galaxy is constructed using the model of \cite{Mo1998},
which has a dark matter halo with a Hernquist density profile \citep{Hernquist1990}, a thin
disk with gas and stars, and a black hole in the center. The galaxy
properties are similar to those of the Milky Way: the total mass is
$10^{12}\, \Msun$, the gas fraction $f_{\rm gas} = 0.2$, the radial scale length of the galaxy is 3~kpc, and the disk height is
one-fifth of it.  The disk contains $4\%$ of its total mass, and the seed mass of the black hole is
$10^{5}\, \Msun$.

The interstellar medium (ISM) in these galaxies is modeled using a
sub-grid multi-phase recipe, and the star formation rate follows the
empirical Schmidt- Kennicutt law \citep{Schmidt1959, Kennicutt1998}
where the surface density of star formation is related to the surface
density of gas ($\Sigma_{\rm SFR} \propto \Sigma_{\rm gas}^{1.4}$).  The
radiative cooling and heating in the ISM is modeled with the
assumption that the medium is in collisional equilibrium and there is
an external UV background \citep{Haardt1996}. We have also followed
feedback processes from both supernovae \citep{Springel2003} and active galactic nuclei
\citep{Springel2005bh}. Supernova
feedback includes thermal energy and galactic winds. The wind energy
efficiency is 5$\%$ of the supernovae energy, and the wind direction
is anisotropic: winds carry energy and matter perpendicular to the
disk plane.  The feedback from the black holes is in the form of
thermal energy deposited isotropically into the surrounding gas.

We note that \cite{Renaud2015} have performed a simulation of Antennae-like merger.
In order to explore a more extreme merger environment, we simulate 
a head-on collision of two Milky Way-sized galaxies, in which the progenitors are initially placed on a parabolic orbit with the inclination 
of both  with respective to the orbital plane as $\theta=0$ and $\phi=0$.
In the simulations, each galaxy is initially started with 82,000 gas
particles, 328,000 star particles and 1,476,860 dark matter particles,
which yield a mass resolution of $5 \times 10^4\, \Msun$ for the gas
and star particles, and $6\times10^5\, \Msun$ for each dark matter
particle.

\subsection{Star Cluster Identification}

Finding groups or structures in a given set of data is a classic
problem in data mining. There are many algorithms for group finding,
which essentially differ in their notion of groups and in their
methods. The two main classes are particle-based and density-based
algorithms. The most widely used method of group finding in astronomy
is the FOF algorithm \citep{Davis1985}. It is a
particle-based algorithm where all particles within a given linking
length are considered as a group. However, there are two significant
downsides of this approach even with an adaptive linking length
\citep{Suginohara1992}: i) if two groups have a linking bridge, they
will be identified as one group; ii) it cannot identify substructures
within a structure.

The other class of group-finding algorithms is density-based, which
identify overdensities in the field as groups \citep{ Warren1992,
  Bertschinger1991, Klypin1997, Gill2004}. These methods do not suffer
from the problems of the FOF methods described above. For this reason,
we adopt the density-based hierarchical group finding algorithm AHF
\citep{Knollmann2009} to identify SCs in the simulations
using the following procedures.

First, AHF divides the simulation box into grid regions. It determines
the density inside each grid cell and compares the density to a
threshold or background density value. If the computed density exceeds
the threshold, it divides the grid into half of its initial size.  It
computes the densities in each of the refined grid cells and again
compares with the threshold. This process goes on recursively until
all the cells in the simulation box have densities less than the
threshold value. Next, it starts from the finest grid and marks
isolated overdense regions as possible clusters. It goes on to the
next coarser level and again identifies possible regions as
clusters. Importantly, it links the possible clusters in finer grids
to their respective coarser parts (linking daughters to parents). This
continues until it has reached the coarsest grid and finally it builds
a tree of clusters with subclusters.

Considering the observed physical properties of young massive
clusters, we impose a few criterions on the groups identified by AHF to
qualify them as SCs. 
Each group should be gravitationally bound and have no substructures
in order to only include individual star clusters. It should contain stars and have at least 4 baryonic particles, which means the minimum cluster 
mass is
$2\times 10^5\,\Msun$.  The upper limit of group mass is set at $10^8\,\Msun$, and the baryonic fraction (mass ratio of baryons to dark matter) of 
each group should be larger than unity in order to distinguish the star clusters from dwarf galaxies that may form 
in our simulations.
Such an approach provides a holistic identification of star
clusters in our study.

\section{Formation of Star Clusters} 
\label{sec:formation}

\subsection{Starbursts in Interacting Galaxies} 

\begin{figure}
 \includegraphics[width=0.5\textwidth]{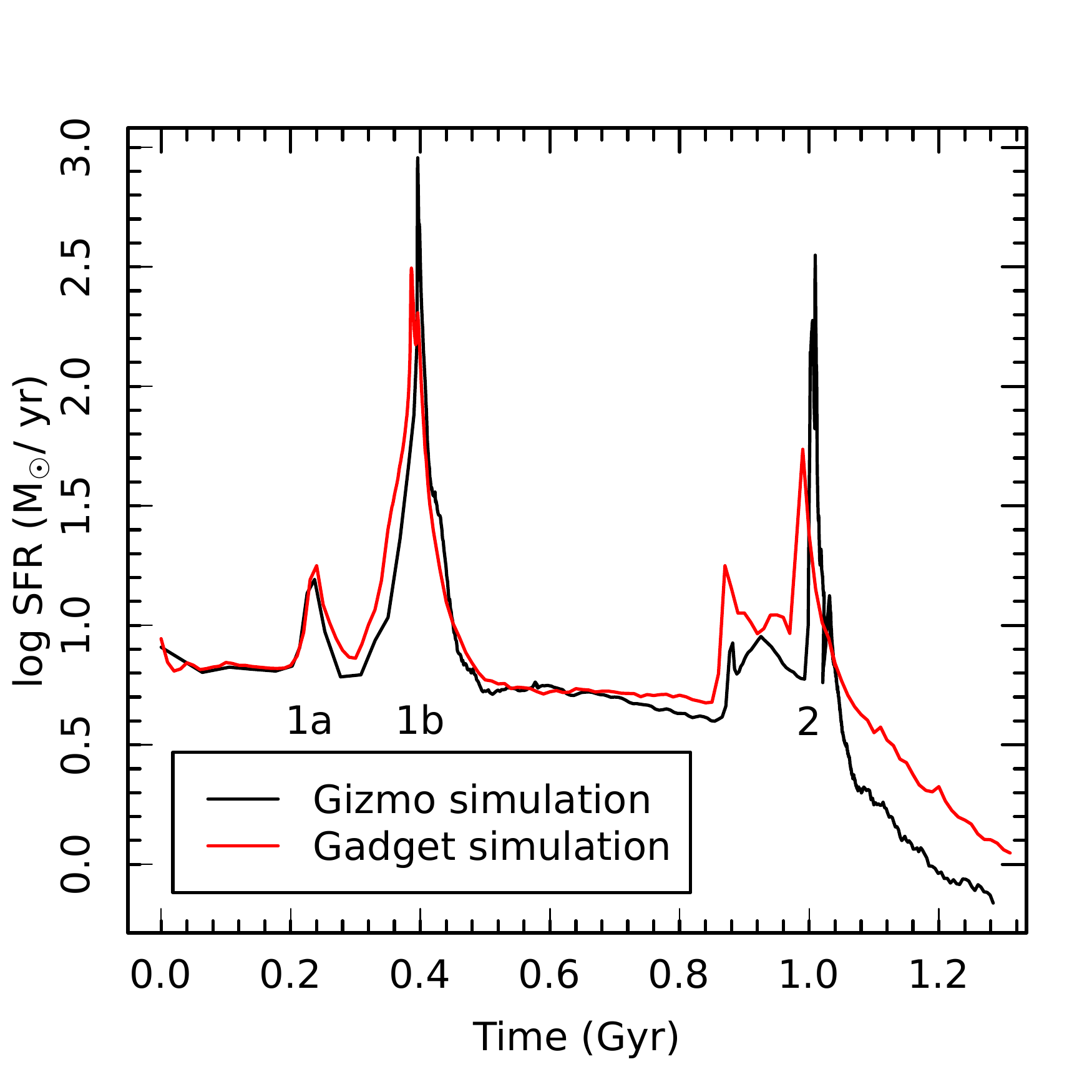} 
 \caption{ The star formation histories of galaxy mergers in both {\Gizmo} (black curve) and {\Gadget} (red curve) simulations. 
Both simulations show two starburst episodes during the close encounters of the two galaxies, at times $\sim 0.2 - 0.45$ Gyr and $\sim 0.8 - 1.1$ Gyr, respectively. In order to investigate the triggering source of star formation, we calculate gravitational torques on star forming gas during three star formation peaks as labeled: minor peak 1a at 0.23 Gyr and major peak 1b at 0.4 Gyr during the first close passage, and major peak  2 at 1 Gyr during the final coalescence, and show the results in Figure~\ref{fig:torque_comp}. }
 \label{fig:SFR}
\end{figure}

 \begin{figure*}
 \centering
  \includegraphics[width=0.32\textwidth]{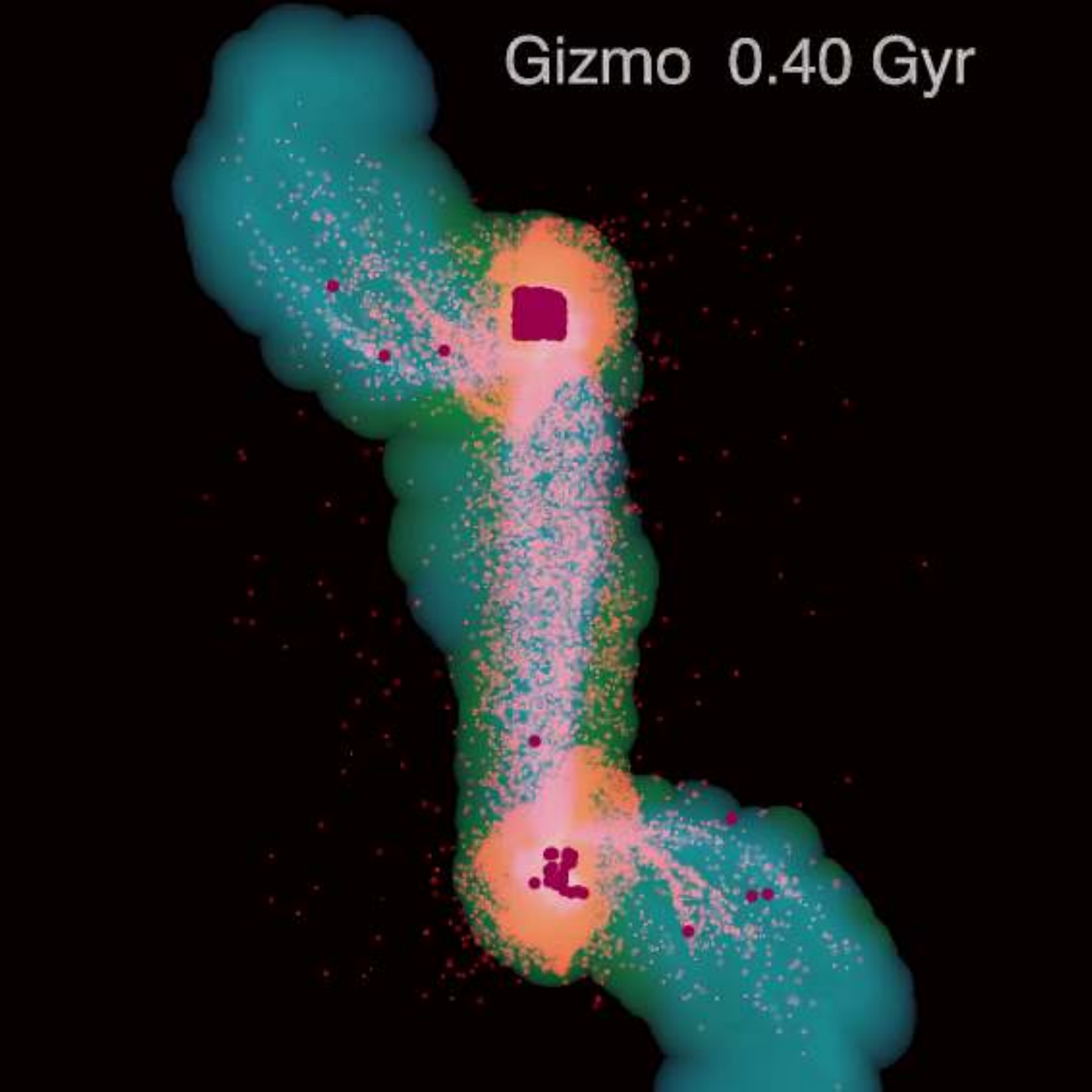}
  \includegraphics[width=0.32\textwidth]{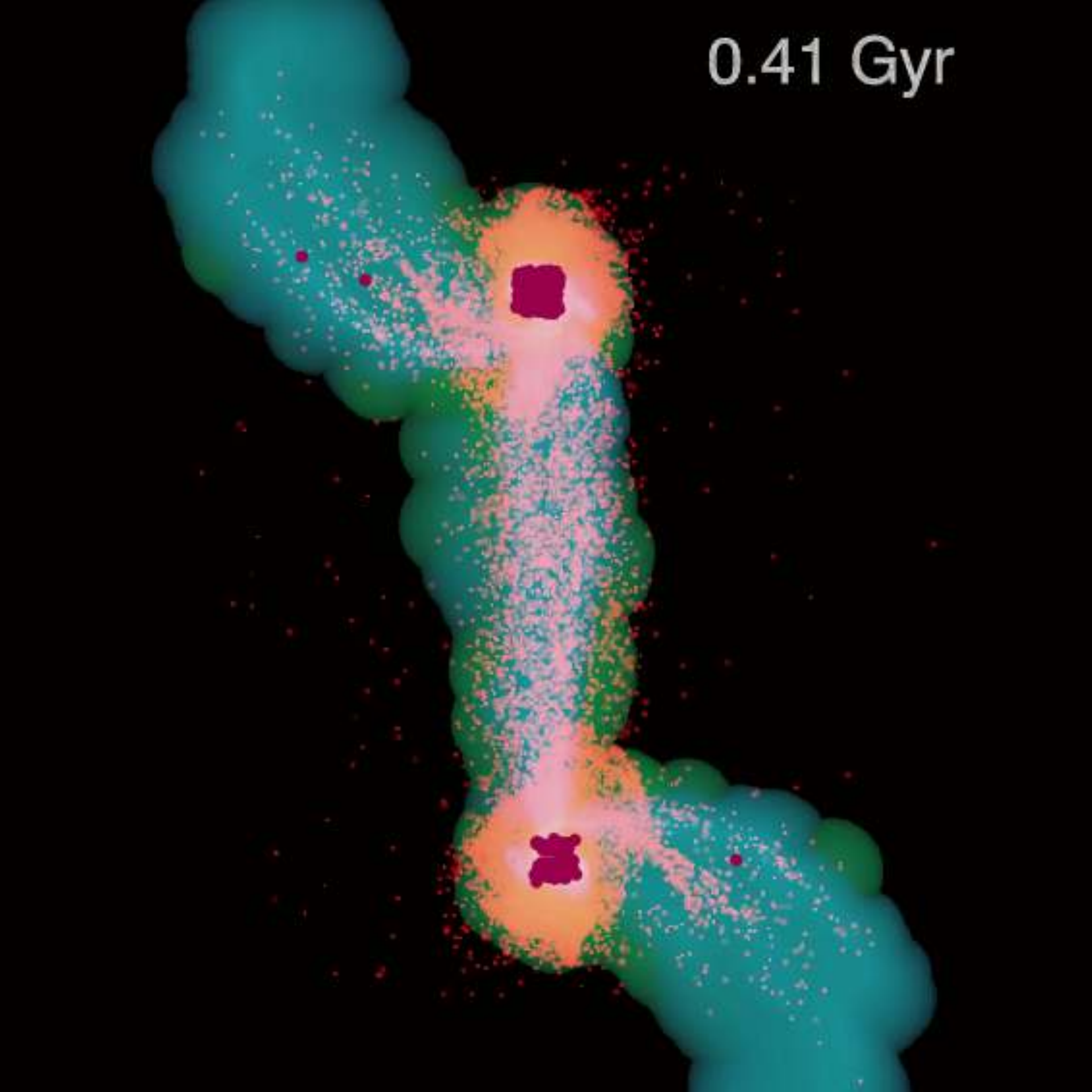}
  \includegraphics[width=0.32\textwidth]{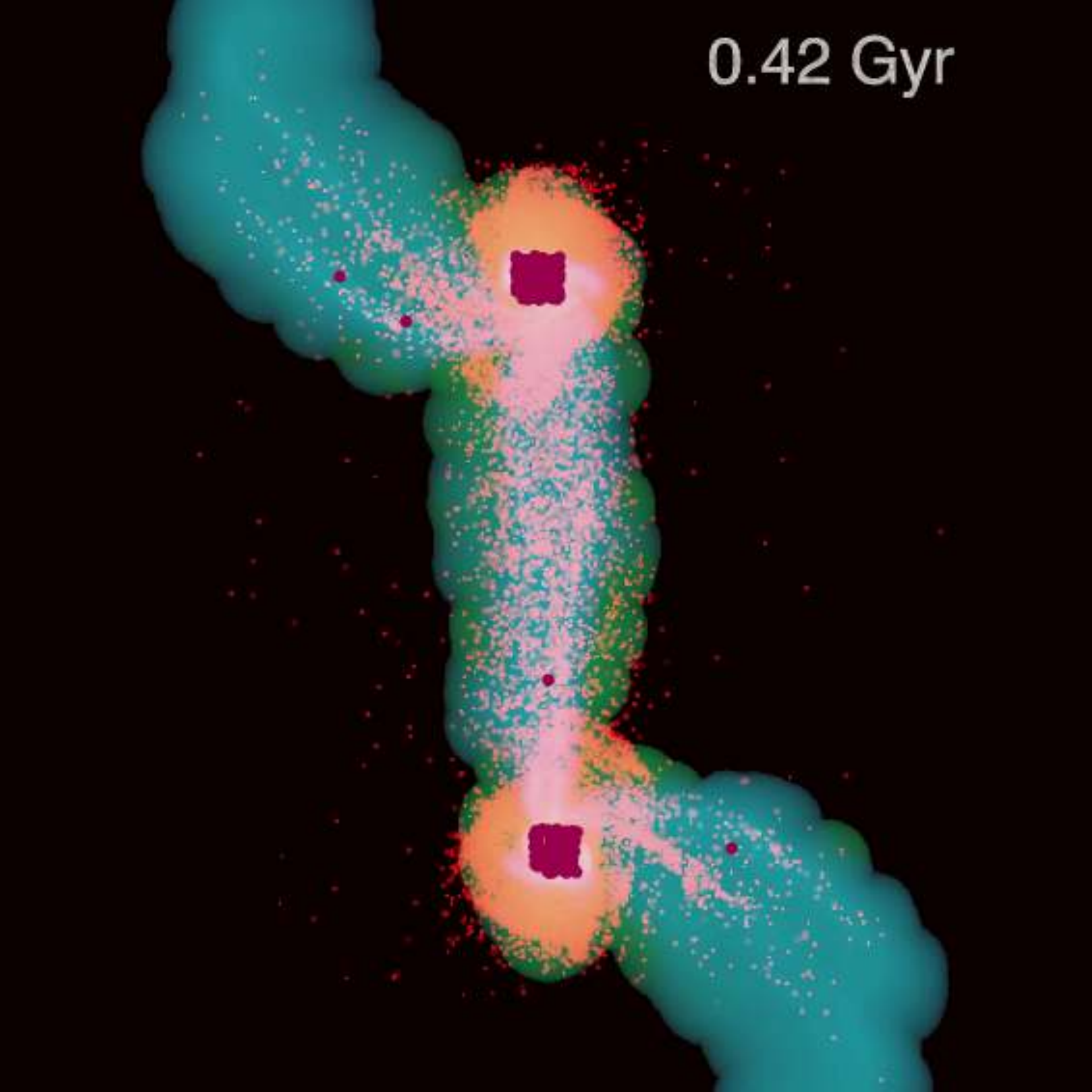}\\
  \includegraphics[width=0.32\textwidth]{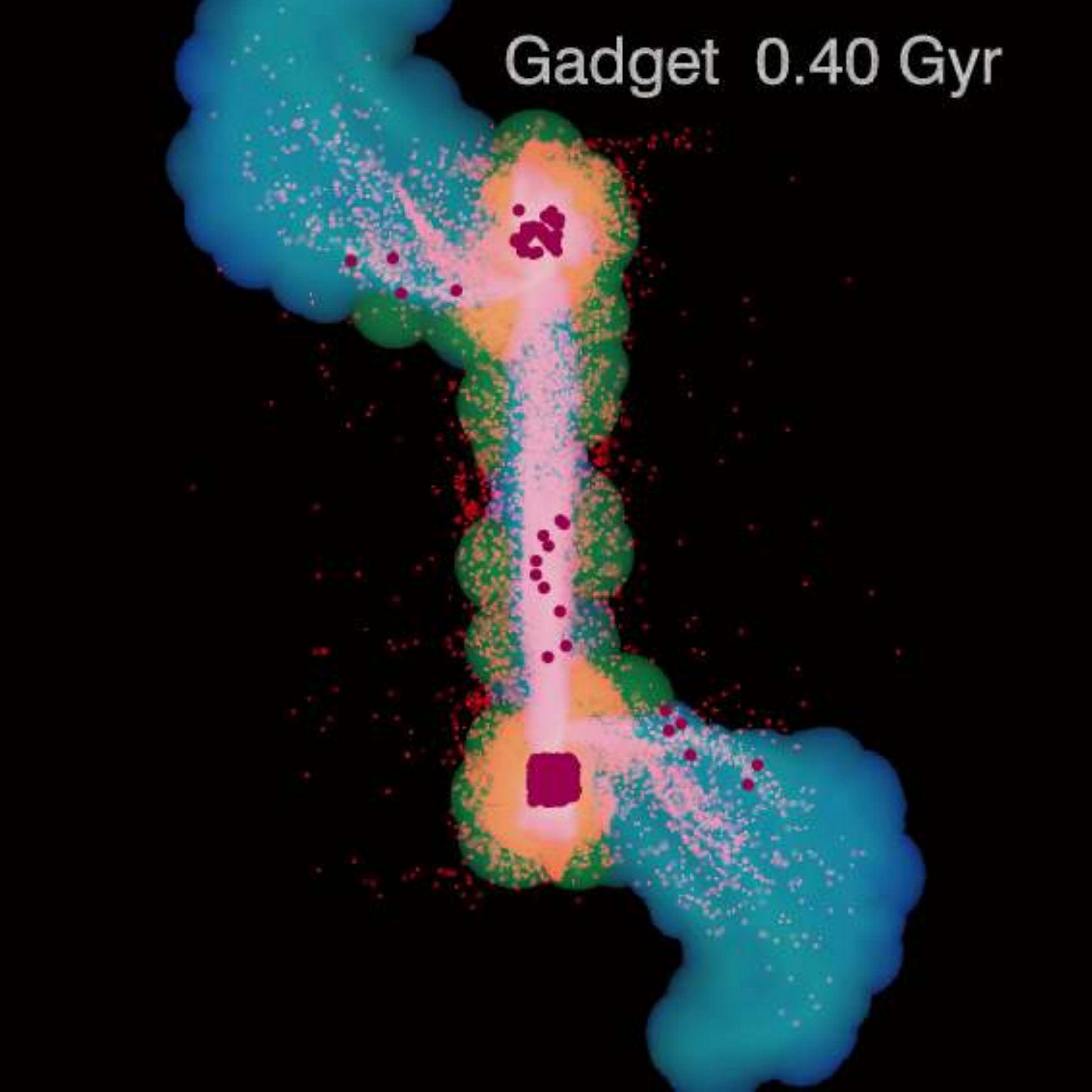}
  \includegraphics[width=0.32\textwidth]{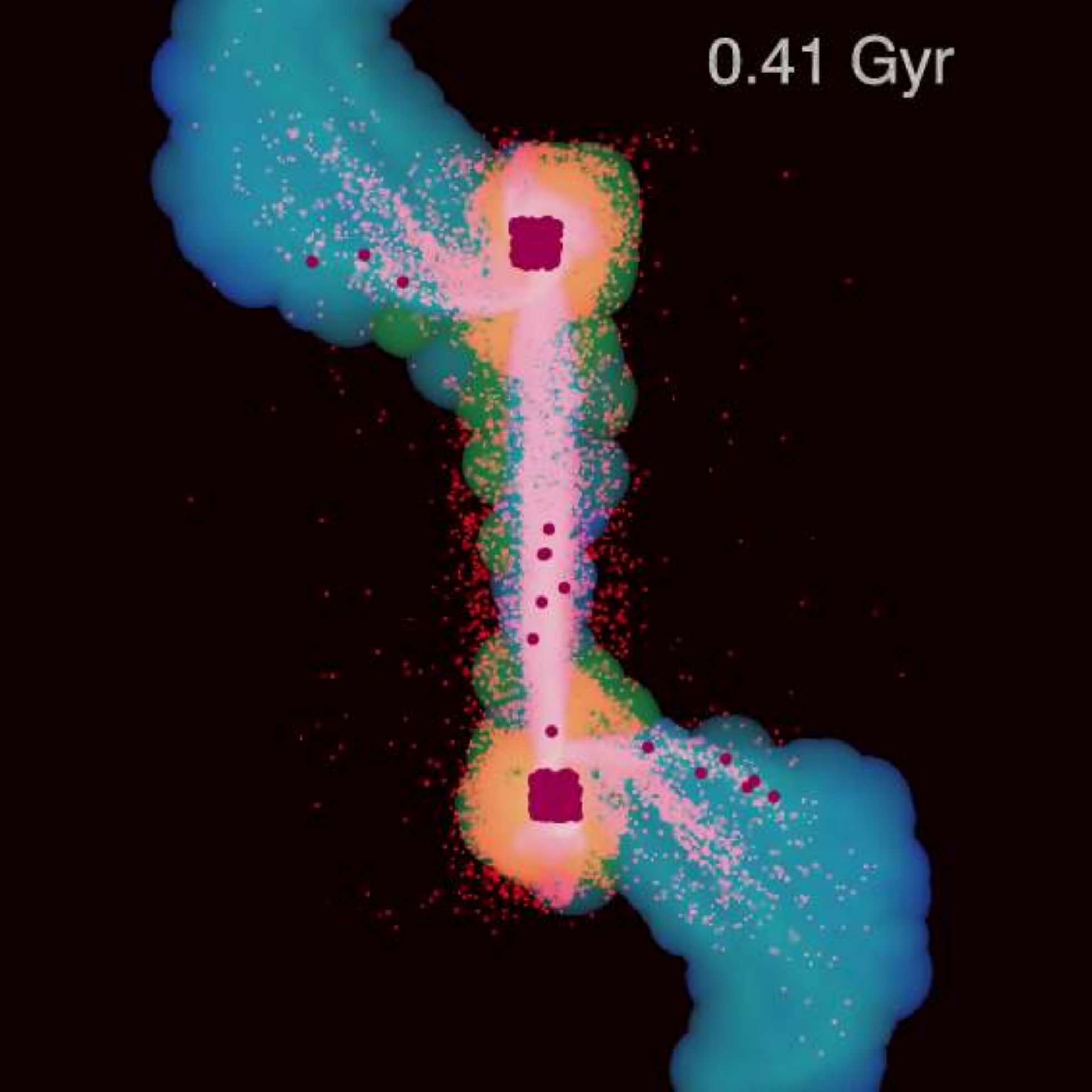}
  \includegraphics[width=0.32\textwidth]{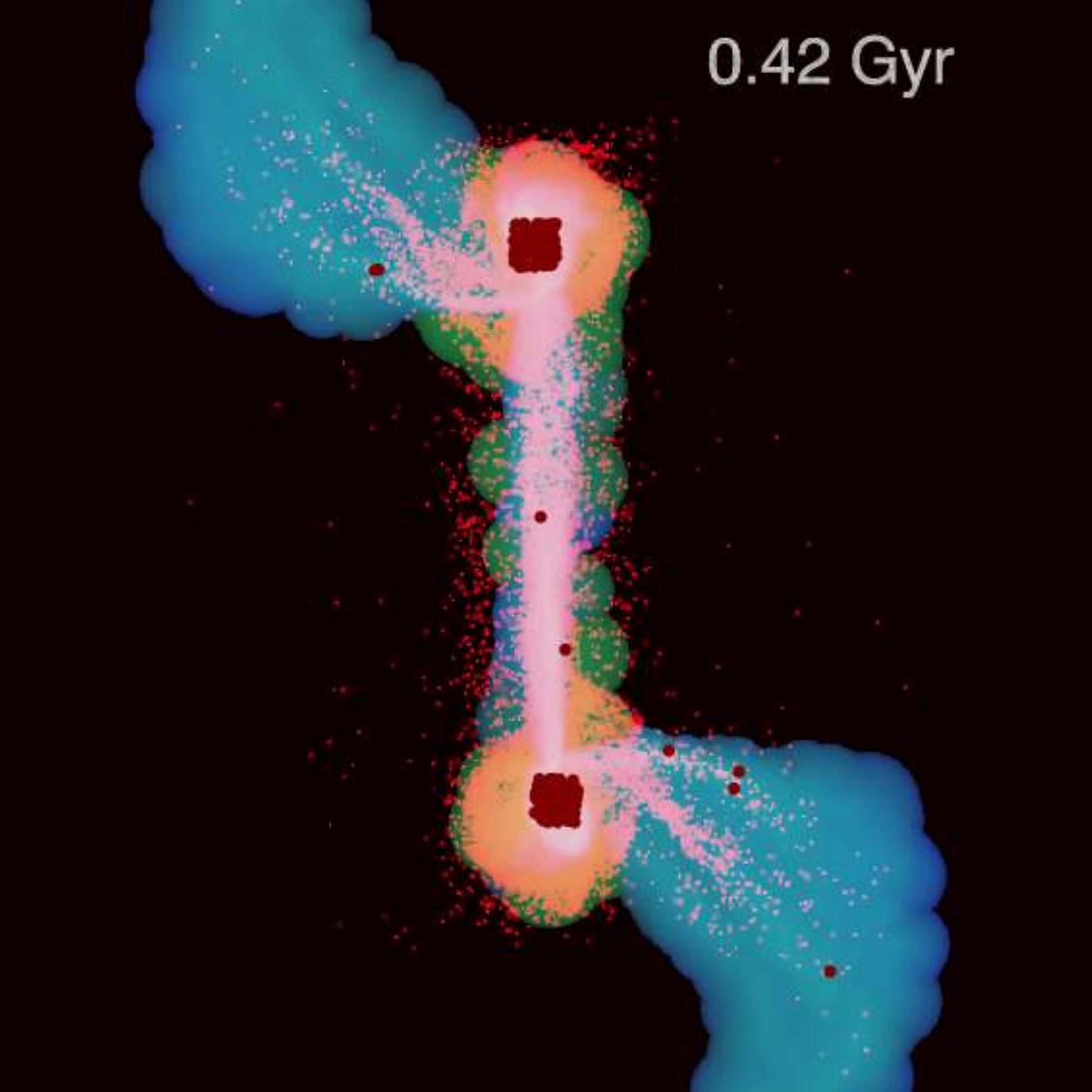}
  \caption{Snapshots of the galaxy merger at three different times, 0.40 Gyr, 0.41 Gyr and 0.42 Gyr during the first starburst phase when  
  most clusters form, from both {\Gizmo} (top panels) and {\Gadget} (bottom panels) simulations. The images are 
  projected gas density maps color-coded by gas temperature (the colors from blue to red indicates hotter gas, the brightness from dark to white 
  measures increasing density). The red dots are stars, and the filled maroon circles represent
  newly formed star clusters. 
 The maroon region in the center of each galaxy indicates overlapping star clusters. The box length is 100 kpc in physical coordinates.}
 \label{fig:snapshots}
 \end{figure*}

In the simulations, the progenitor galaxies start moving towards each
other at $t=0$ on a parabolic orbit. They have their first close
encounter at $t \sim 0.23$ Gyr, and the second one at $t \sim 0.9$ Gyr
until the final coalesce at $t \sim 1$ Gyr. During the close passages,
vigorous star formation is triggered by the compression of gas by
tidal forces and rising gas densities in the inner region of the
galaxies due to gravitational torques \citep{Barnes1991,
  Barnes1996}. As shown in Figure~\ref{fig:SFR}, the first starburst
occurs at $t \sim 0.2 - 0.5$ Gyr, when the star formation rate (SFR)
increases by nearly two orders of magnitude and reaches a peak of
$\sim \rm{ 10^3\, \Msun/yr}$ at $t \sim 0.4$ Gyr. 
The second starburst
takes place at $t \sim 0.9 - 1.1$ Gyr, and the SFR peaks at $\sim \rm{
10^2\, \Msun/yr}$ at $t \sim 1$ Gyr.

The star formation history of galaxy mergers depends strongly on the progenitor properties and orbital parameters. 
The SFR in our simulation is higher than typical mergers at the local universe.
\cite{Renaud2015} estimated from their simulation that the SFR of Antennae merger at its starburst phase 
is $\sim 10^2\Msun$ .
However, star formation in
ultra-luminous infrared galaxies (ULIRGs) in nearby universe can have comparable intensity. For example, 
radio recombination line studies of merer driven starburst galaxy Arp 220 (77 Mpc away) suggest a 
mean SFR of $\sim 240\, \Msun/yr$ or more plausibly short periods of intense starbursts with SFR of $\sim 10^3\, \Msun/yr$ 
\citep{Anantharamaiah2000,Thrall2008,Varenius2016}. 
We note that the mass of Arp220 is estimated as  $\sim 10^{10}\,\Msun$ \citep{Scoville1997}, much lower than our modeled
galaxies. The high SFR in our simulated galaxies may be a product of both their high mass progenitors and their
extreme orbital parameters with the head-on collision.

As demonstrated in Figure~\ref{fig:SFR}, there is a remarkable difference in the star formation histories
between the two simulations in that {\Gizmo} produces higher SFR peaks
than {\Gadget} by a factor of $3-5$. This is due to the more accurate
treatments of fluids and shocks in {\Gizmo}. Similar differences have
also been seen in the code comparison study of galaxy mergers using
{\Gadget} and the moving-mesh code {\Arepo} by \cite{Hayward2014}, who
reported that {\Arepo} produces higher SFRs than {\Gadget} by up to a
factor of 10 for mergers of Milky Way - size galaxies.

The strong compression and shocks produced by the galaxy interaction
fuel rapid formation of SCs during the starbursts. As
demonstrated in Figure~\ref{fig:snapshots}, most of the SCs
form in the nuclear regions of the two merging galaxies, with a few
spread in the tidal tails and the galactic bridges. Similar
distributions of YMCs have also been observed in
galaxy mergers, including nuclear region clusters by
\cite{Whitmore1995, Miller1997}, and tidal tail clusters by
\cite{Barnes1992, Knierman2003, Bastian2005, Mullan2011}.

\begin{figure}
\centering
\includegraphics[width=0.45\textwidth]{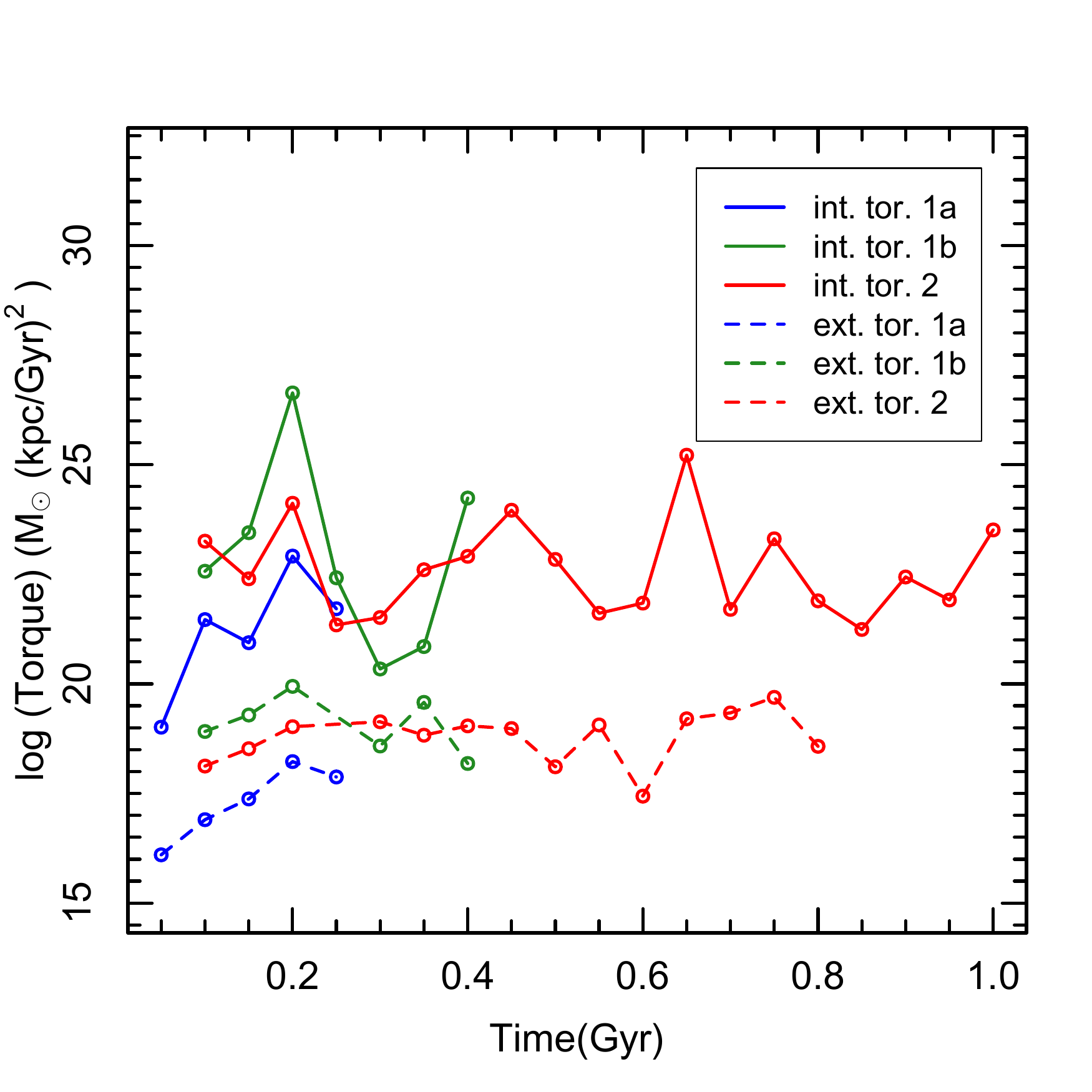} 
\caption{Evolution of the gravitational torques on  the gas particles that 
eventually form stars during the three star formation peaks, 1a (at time 0.23 Gyr, \textit{blue}),  1b (at 0.4 Gyr, \textit{green}), and 2 (at 1 Gyr \textit{red}), 
as labeled in Figure~\ref{fig:SFR}. The solid and dashed lines represent internal and external torques, respectively. Note that during the final coalescence at time $\sim 0.8 - 1.1$ Gyr, only internal torque is available.} 
\label{fig:torque_comp}
\end{figure}

In order to investigate the triggering source of star formation during the merging process, we track the gas particles that form stars at the three star formation peaks, minor peak 1a at 0.23 Gyr and major peak 1b at 0.4 Gyr during the first close passage, and major peak  2 at 1 Gyr during the final coalescence, as labeled in Figure~\ref{fig:SFR}. It was shown by \cite{Hernquistnature1989} that the major star formation episodes in galaxy mergers are marked by a rapid loss of the angular momentum of the star forming gas driven by the gravitational torque. We follow the procedure of \cite{Barnes1996} and calculate the gravitational torque, $\tau = r \times F$, exerted on these star-forming gas particles by the 
gas and stars in the same galaxy (internal torque), and by gas, stars and halo particles of the other galaxy (external torque). As shown in Figure~\ref{fig:torque_comp}, the internal torque is  higher than the external counterpart by orders of magnitude for all tracked star-forming gas particles during the galaxy interaction. Similar results have been reported by a number of theoretical studies of major mergers ( e.g. \citealt{Hernquistnature1989, Mihos1994, Barnes1996, Hopkins2009}), which show that the internal torque is the dominant source of torque that drives the loss of angular momentum in these interacting galaxies. The close encounter of the galaxies produces strong tidal forces that results in a non-axisymmetric response in the galaxy disks. These forces deform the galaxy disks and form gaseous and stellar bars in the galaxies. These gas bars lead the stellar bars by a few degrees \citep{Barnes1991} which eventually produces
a strong torque on the gas near the center that drives rapid gas inflow towards the nuclear region, resulting in vigorous starburst.  From Figure~\ref{fig:snapshots}, the majority of clusters   formed at the starburst phase are highly clustered in the center region of each galaxy, indicating their origin from the nuclear gas inflow, while the few clusters formed in the bridge and tidal tails may be triggered by tidal force, as suggested by \cite{Renaud2009} and \cite{Renaud2014}. 

We note that the first major star formation peak 1b (at 0.4 Gyr) takes place about 160 Myr after the first close pericentric passage at $\sim 0.24$ Gyr. Similar time delay has been found in other simulations of galaxy mergers (e.g., \citealt{Mihos1994, Mihos1996, Cox2008, Hayward2014}), as the timescale to build up the gas density driven by internal torque for star formation. 
In addition, we note that there is a minor star formation peak of $\sim 15\, \Msun/yr$ at 1a (0.24 Gyr) preceding the the major one of $\sim 800\, \Msun/yr$ at 1b (0.4 Gyr). We find that the ratio of external to internal torque peaks during the 1a phase, suggesting that external torque from tidal force may contribute to the star formation as well. Studies by \cite{Renaud2009, Renaud2014} have shown that tidal force during galaxy interaction may compress the gas and enhance the star formation.

\subsection{Initial Cluster Mass Functions}

 \begin{figure}
 \centering
 \includegraphics[width=0.4\textwidth]{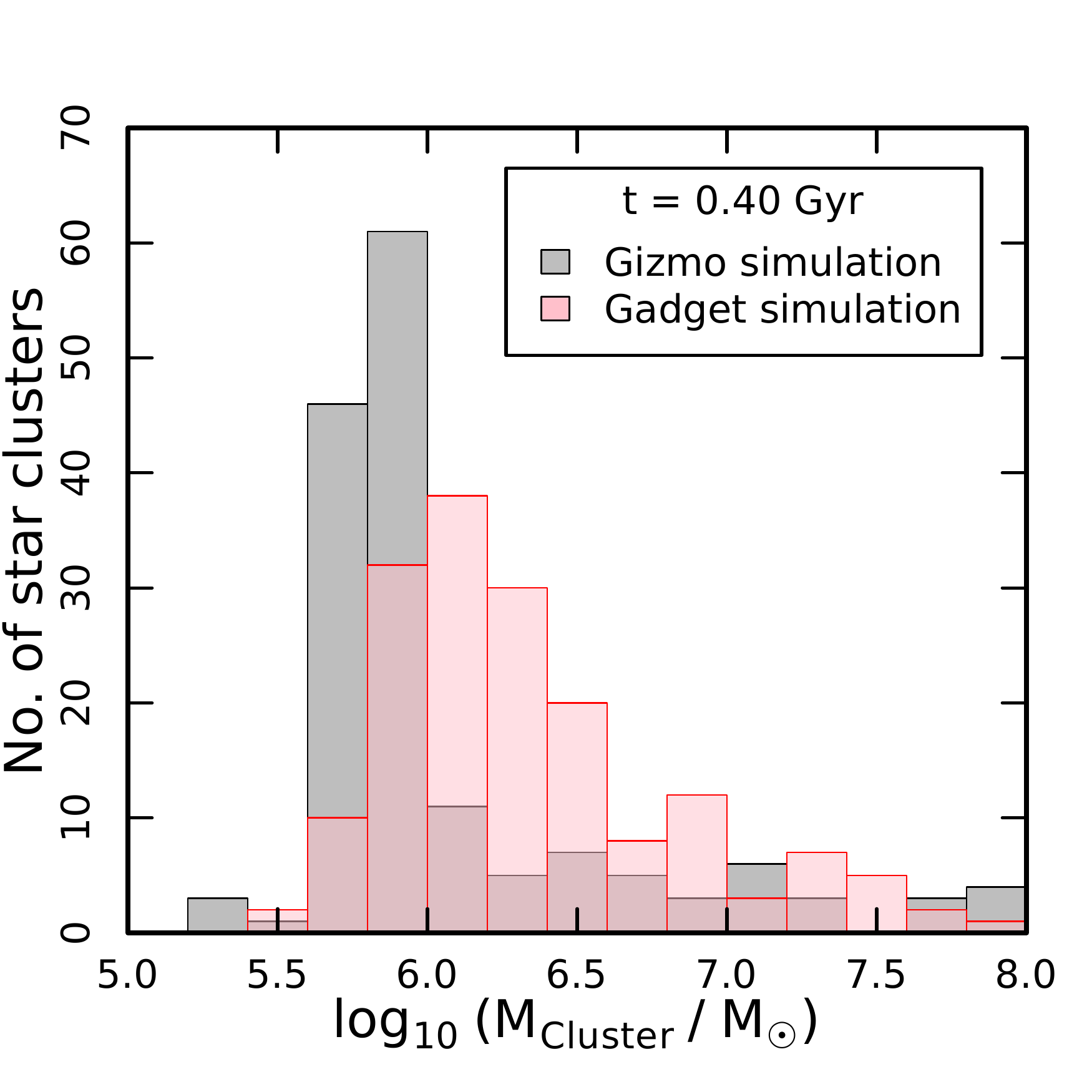}
 \includegraphics[width=0.4\textwidth]{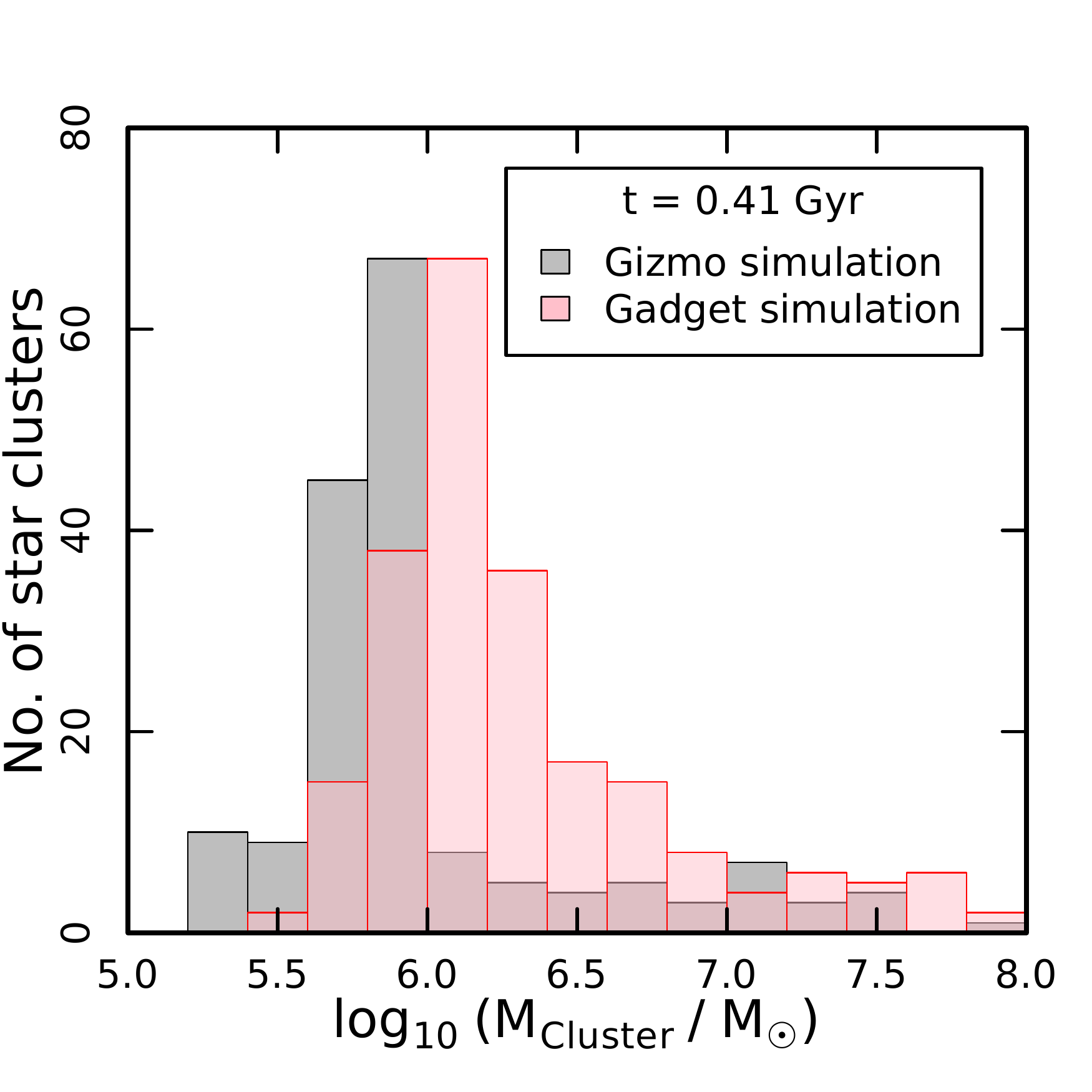}
  \includegraphics[width=0.4\textwidth]{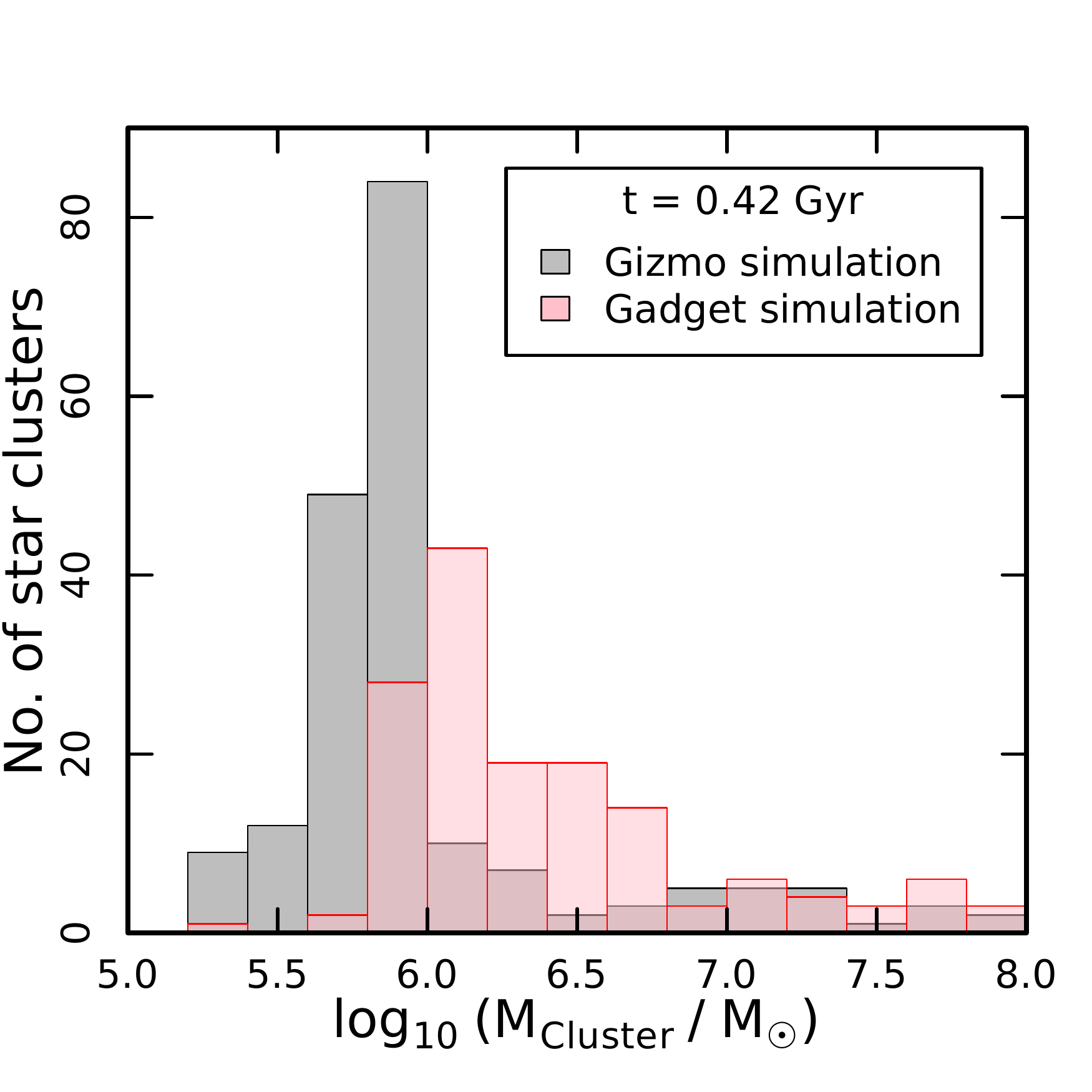}
\caption{Mass functions of star clusters formed at 0.40 Gyr, 0.41 Gyr and 0.42 Gyr during the 
first starburst phase when  
  most clusters form, from both {\Gizmo} (grey) and 
{\Gadget} (red) simulations. The total number of clusters in the three snapshots 
from {\Gizmo} ({\Gadget}) simulation is 158 (170), 171
(221), and  197 (151), respectively.}  
  \label{fig:icmf}
\end{figure}

The resulting mass functions of the SCs formed during the
first close encounter are shown in Figure~\ref{fig:icmf}. Although the
total number of clusters in the same snapshot differs between the
{\Gizmo} and {\Gadget} simulations by a factor of $\sim 1.3$, the
range of $\sim 150 - 200$ is in good agreement with observations of
galaxy mergers such as the Antennae \citep{Whitmore1999,
  Larsen2010}. More interestingly, both simulations produce similar
mass distributions which resemble a peaked or a quasi-lognormal
function, with a {\Gizmo} peak around $10^{5.8 - 6}\, \Msun$ and the
{\Gadget} peak at $10^{6 - 6.2}\, \Msun$.

Our mass functions do not show a purely declining power law, as
suggested by many observations of YMCS \citep[e.g.,][]{Zhang1999,
  BiK2003, McCrady2007, Fall2012}. This could be due to the limited
resolution in our simulation so we cannot resolve clusters at mass lower than $10^5\, \Msun$. 
 However, \cite{Renaud2015} also reported lognormal-shape
ICMFs from their Antennae simulation even though they have a much
higher mass resolution ($\sim 70\, \Msun$). It is also possible that
the power-law phase is extremely short lived ($< 10$ Myr) in violent
mergers and our snapshot interval (every 10 Myr) misses that phase. We
note, however, some observations have suggested that the power law
ICMF for YMCs is not universal, rather they have a turnover at low
mass \citep{Cresci2005, Anders2007}.

The peak masses of these ICMFs are higher than those of the observed GCMFs,   
$\sim 1.5 - 3\times 10^5\,\Msun$, but it can be predicted that 
after an evolution of Gyrs, stellar evolution, tidal stripping and 
other disruptive processes will cause them to lose some of their mass
\citep{Kruijssen2015, Webb2015}. We discuss their evolution in more
detail in \S\ref{sec:evolution}.
It is encouraging that both of our simulations, together with that by
\cite{Renaud2015}, produce similar quasi-lognormal ICMFs with similar
peak positions, despite having vastly different hydrodynamic solvers,
feedback processes and numerical resolutions. This agreement suggests
that the lognormal mass function is unlikely due to numerical
artifacts but has a physical origin, which will be investigated in the
next section.

\subsection{Physical Conditions of Cluster Formation and Origin of Lognormal Cluster Mass Functions}
 
\begin{figure*}
\centering
\includegraphics[width=0.39\textwidth]{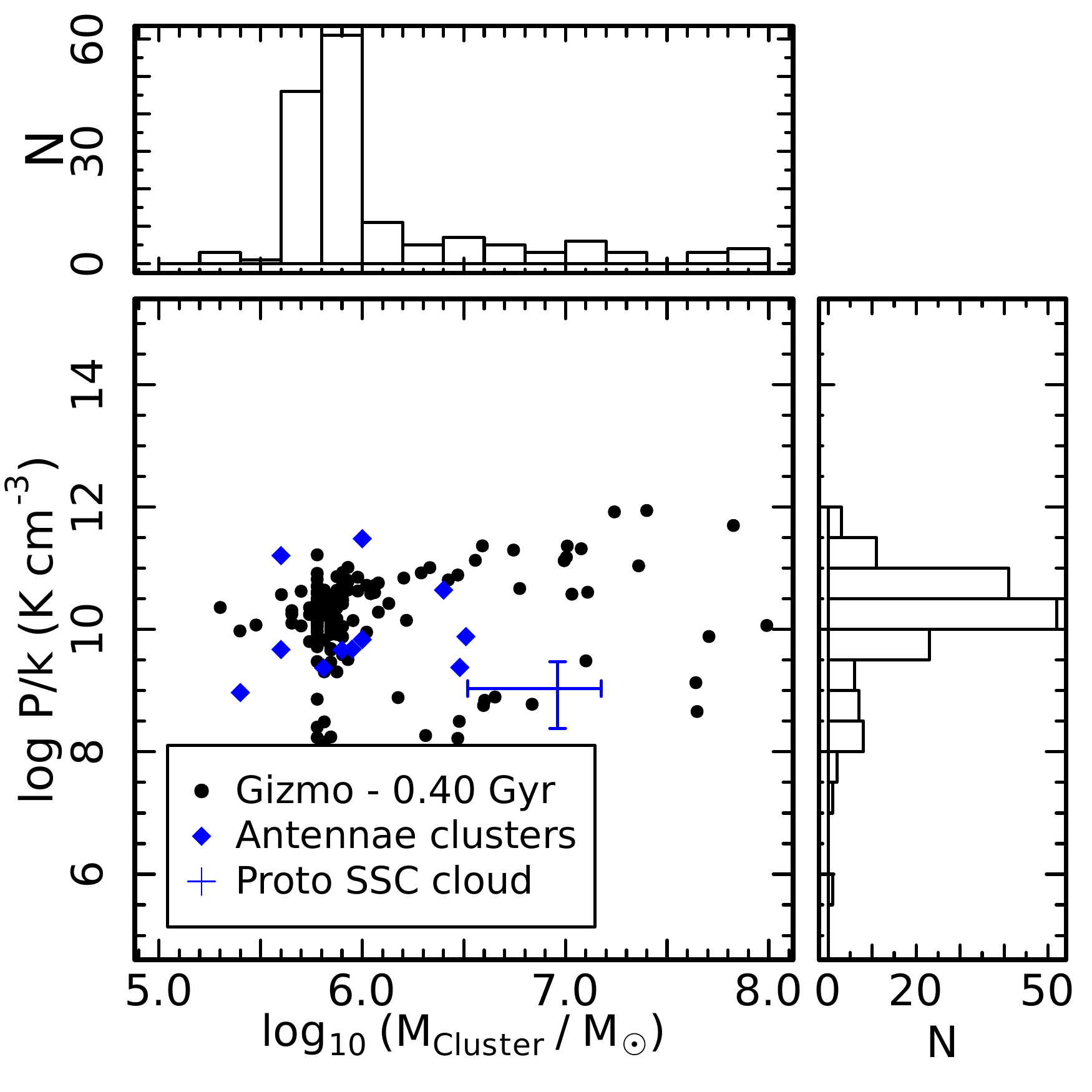}
\includegraphics[width=0.39\textwidth]{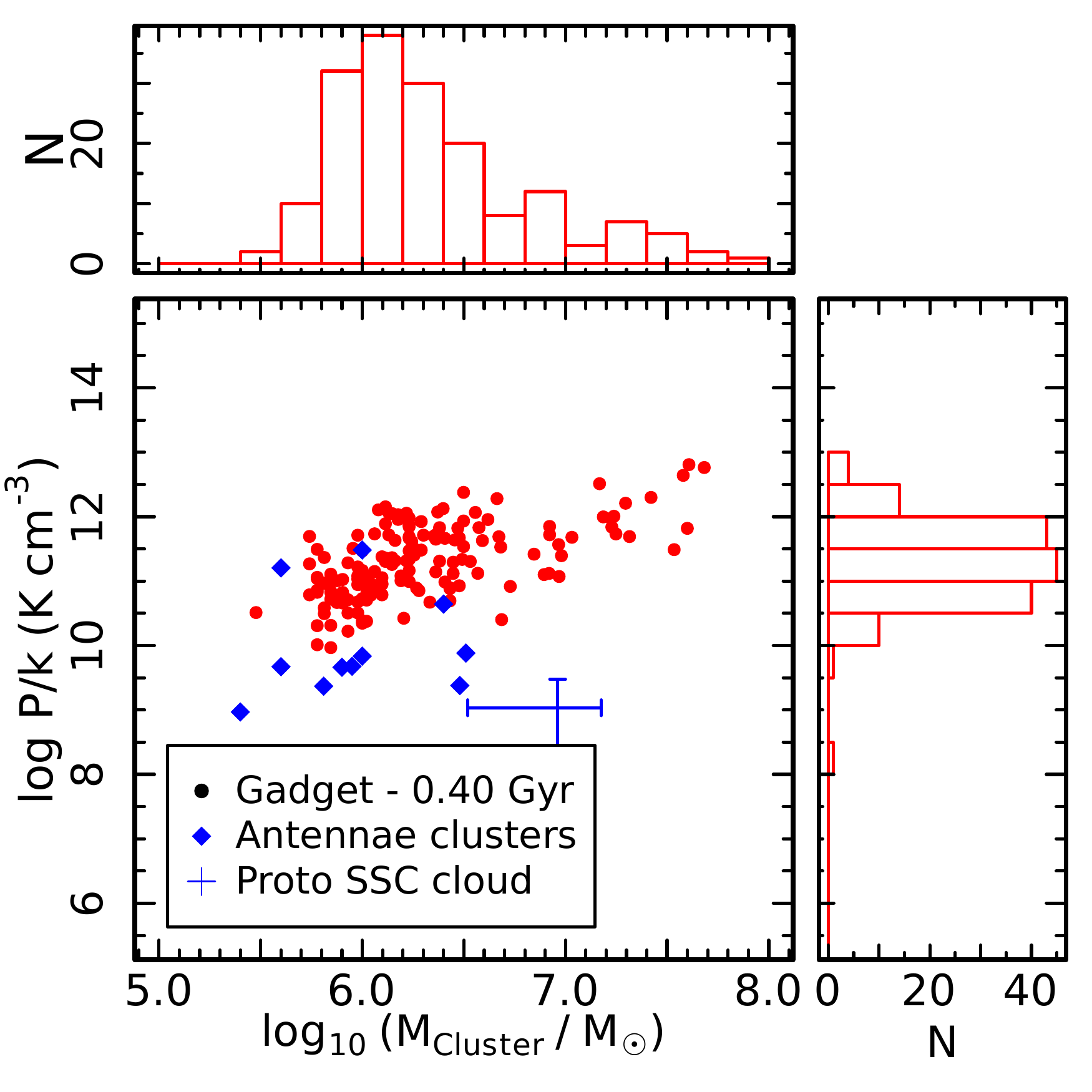}\\
\includegraphics[width=0.39\textwidth]{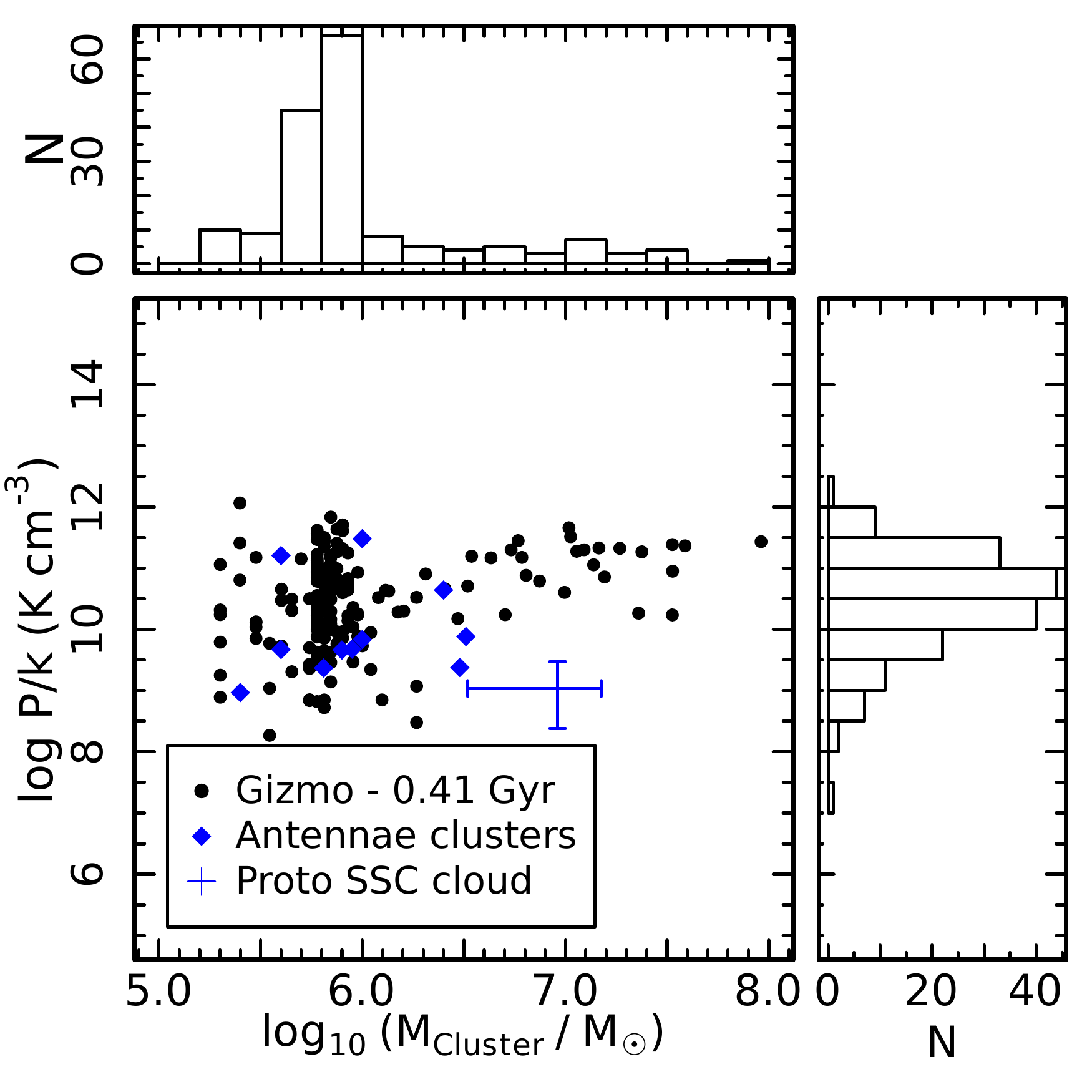}
\includegraphics[width=0.39\textwidth]{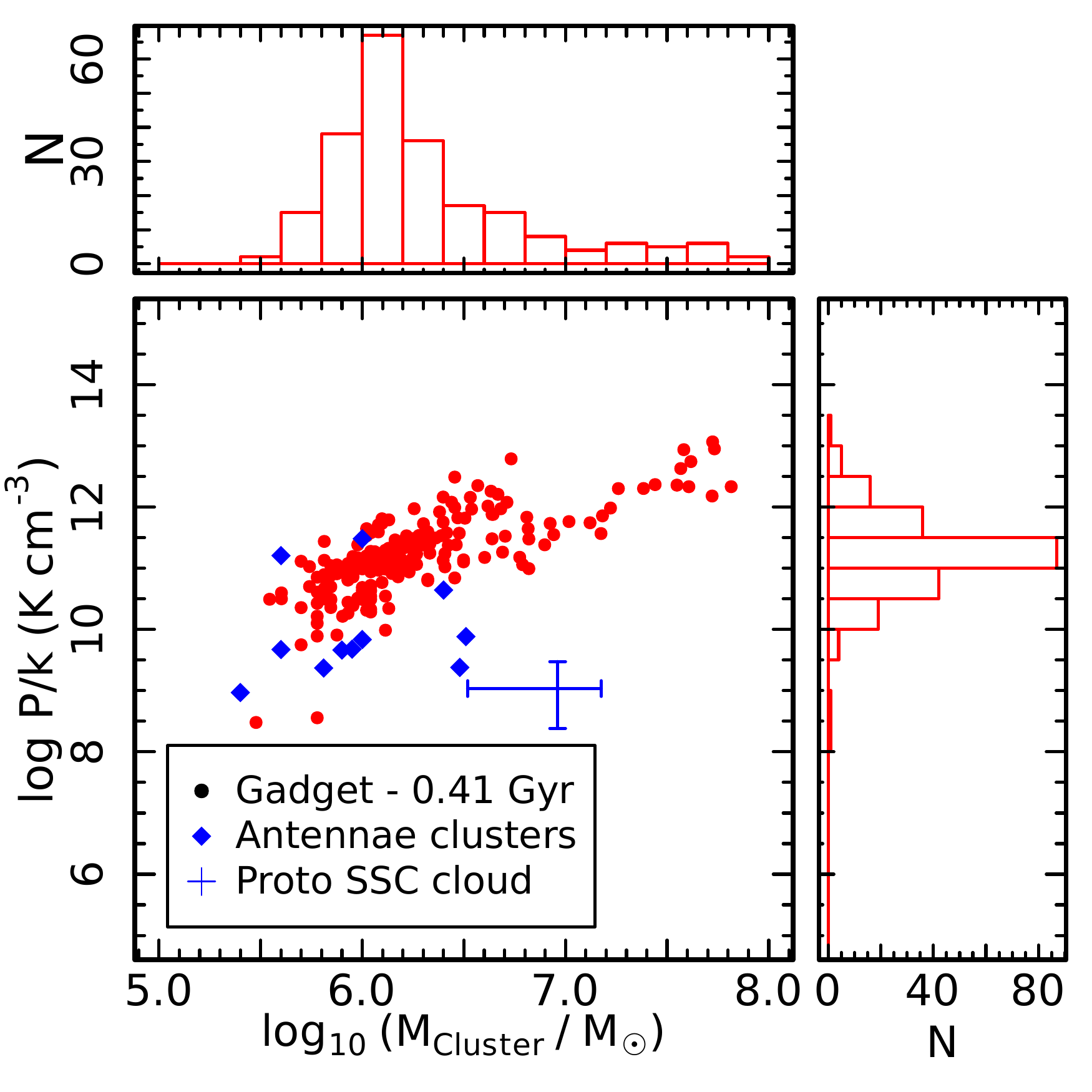}\\
\includegraphics[width=0.39\textwidth]{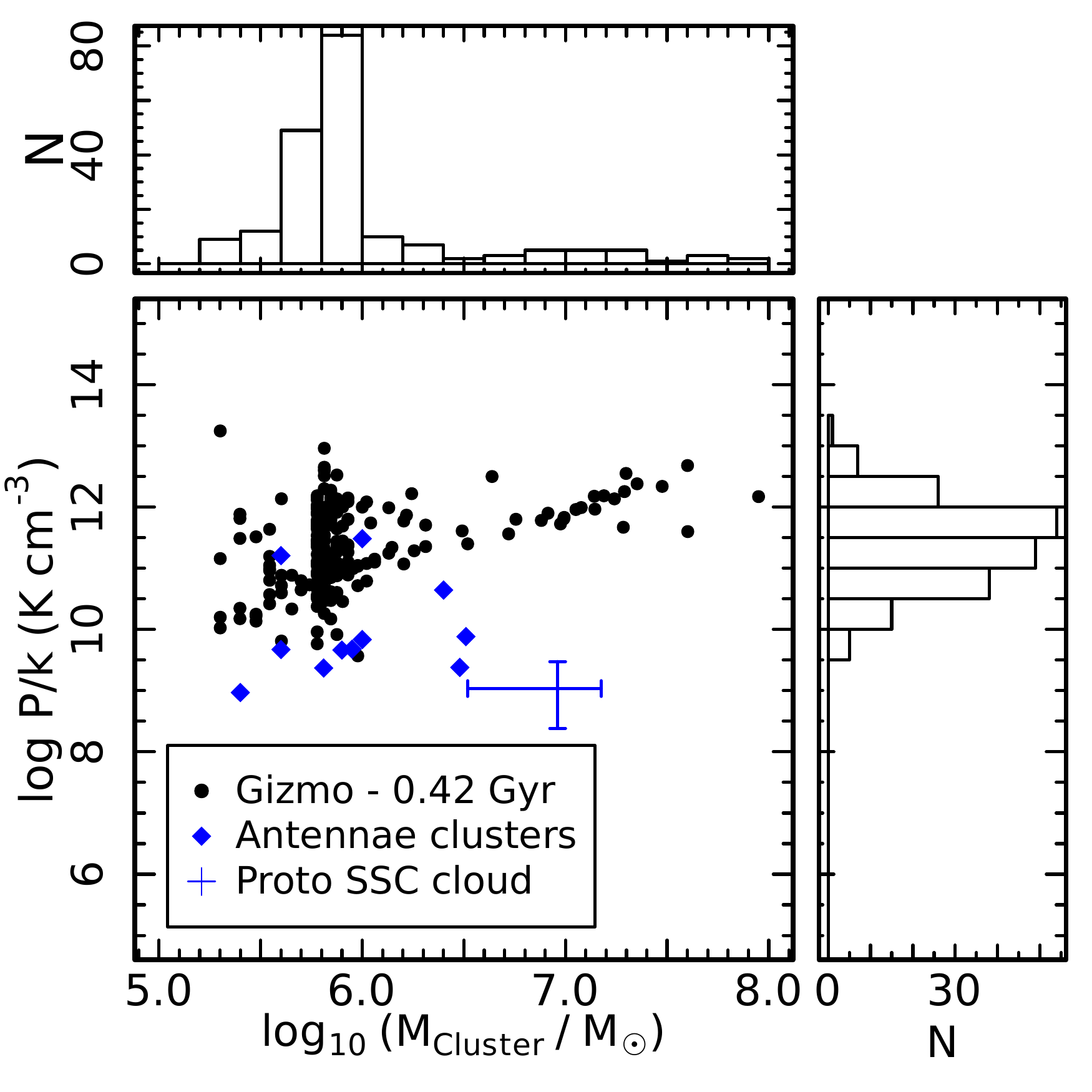} 
\includegraphics[width=0.39\textwidth]{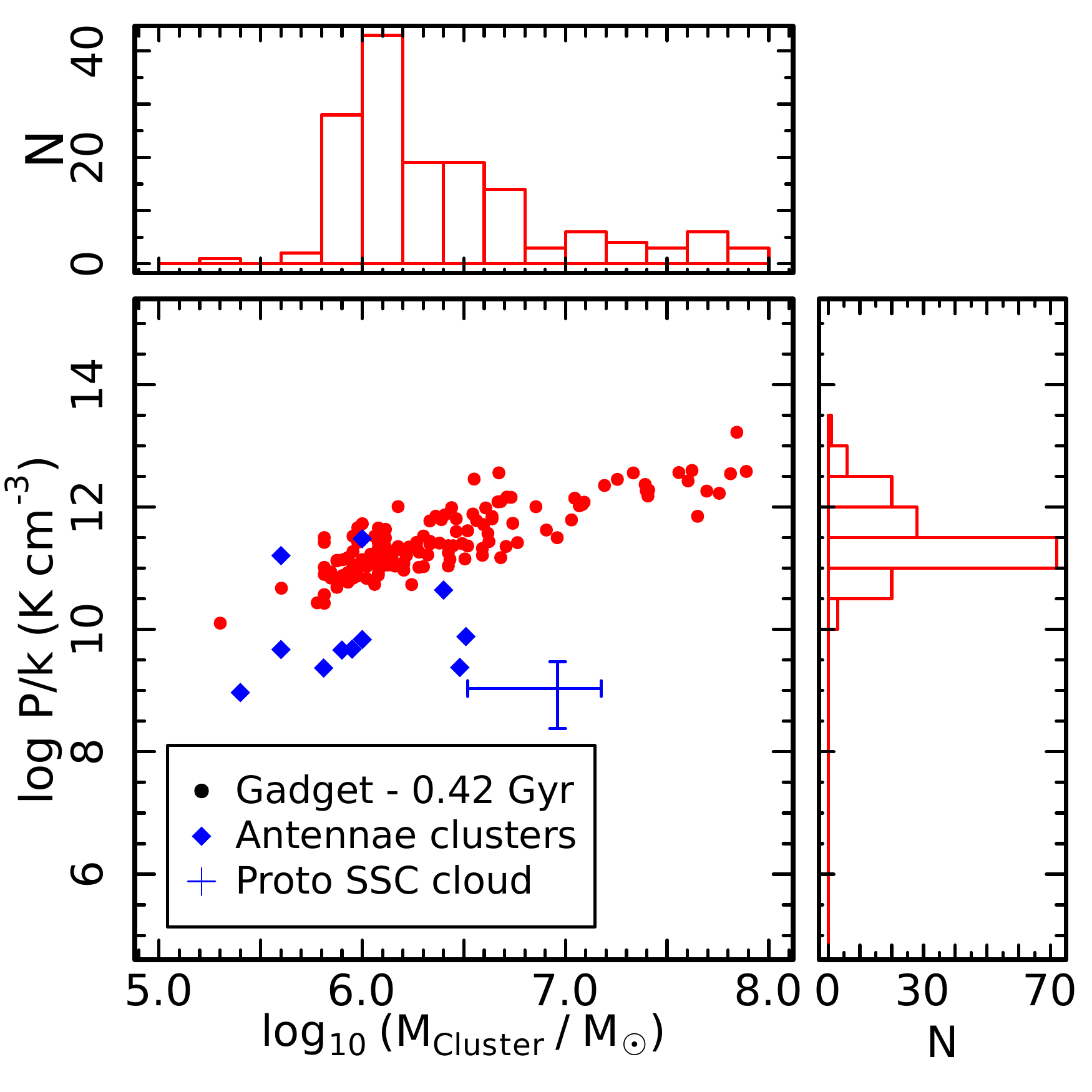}
\caption{Correlation between pre-cluster gas pressure distributions and initial cluster mass functions at 0.40 Gyr, 
0.41 Gyr and 0.42 Gyr during 
the first starburst phase, when  
most clusters form, from both {\Gizmo} (black) and {\Gadget} (red) simulations. The pressure derived from the observed 
pre-super
star cluster cloud in the Antennae by \cite{Johnson2015} is represented by the blue cross where the error bars reflect 
the observational 
uncertainties in cloud mass ($3.3 - 15\times 10^6\, \Msun$), radius ($24\pm 3$ pc) and velocity dispersion ($49\pm 3$ 
km/s).
The pressures derived from observed star clusters in the Antennae
using velocity and radius data compiled in \cite{Zwart2010} and presented
by \cite{Mengel2002, Mengel2008} are represented by blue diamonds. The pressure is expressed as $P/k$ where $k$ is the 
Boltzman constant. } 
\label{fig:pressure}
\end{figure*}

In order to explore the physical origin of the quasi-lognormal ICMFs
in Figure~\ref{fig:icmf}, we examine the
physical conditions of cluster formation. As mentioned in
\S~\ref{sec:intro}, massive SCs form in molecular clouds
with very high gas pressures \citep{Elmegreen1997, Ashman2001}. When a
star-forming cloud is under high pressure, the efficiency of star
formation increases and the gas velocity dispersion becomes higher,
which in turn help to keep the cloud bound \citep{Jog1992}. These
physical conditions create an ideal nursery to form massive,
gravitationally-bound clusters. Recently, \cite{Zubovas2014} performed
an N-body simulation of a molecular cloud and concluded that high
external pressures drive efficient star formation and can cause
cloud fragmentation, leading to the formation of star
clusters. Quantitatively, the external pressure $P$ on a nascent
molecular cloud is given by \cite{Elmegreen1989}:

\begin{equation}
 P = \frac{3\Pi M_{\rm cloud}\sigma_v^2}{4\pi r^3} 
 \label{eqn:1}
\end{equation}

where $M_{\rm cloud}$ is the mass of the cloud, $\sigma_v$ is the velocity
dispersion and $r$ is the size of the cloud.  The factor $\Pi$ is
given by the ratio of density at the cloud edge and the average
density, $\Pi = n_e/\langle n_e\rangle$. 
This $\Pi$ ratio is dependent on the density profile of the parent molecular cloud.
Locally the probability distribution function of ISM density due to turbulence can
be approximated as lognormal but 
at high density regions ($> 10^3\, \rm{cm}^{-3}$), such as at the center of molecular cloud,
density profile can develop a power law tail. At these dense places the $\Pi$ value can be $>> 1$
and consequently it can increase the amount of gas above a density threshold which facilitates 
further star formation \citep{Elmegreen2011, Renaud2014}.

\cite{Elmegreen1997} estimated that the pressure in the birth
clouds of typical GC progenitors or SSCs is $\gtrsim
10^8\, \rm{K}\,\rm{cm}^{-3}$, which is $\gtrsim 10^4$ times
higher than the ISM pressure in the Milky Way \citep{Jenkins1983, Boulares1990, Welty2016}.

In the simulations, in order to determine the pressure of the clouds
from which the SCs form, we track the cluster members back
in time.  We take the constituent star particles of a cluster and
identify the gas particles from which they formed. We then measure the
velocity dispersion of these gas particles and approximate the gas
cloud radius as the average distance from gas particles to the center
of mass of the cloud. 
Due to the limited spatial and mass resolutions, we cannot directly probe the cloud
density profile for estimating $\Pi$, so we approximate $\Pi = 0.5$ following
\cite{Johnson2015}. We note that realistically $\Pi$ can be higher, but it would increase
all our pressure measurements similarly.
With these parameters, we can then calculate the
cloud pressure using Equation~\ref{eqn:1}. In Figure~\ref{fig:pressure}, we
show the resulting pressure distributions against the cluster mass
distributions, in comparison with the pressure of a pre-SSC 
cloud observed in the Antennae by \cite{Johnson2015}. We also
calculate the pressures of observed SCs in the Antennae
Galaxies using the velocity and radius data compiled in
\cite{Zwart2010} and \cite{Mengel2002, Mengel2008} for comparison.

As shown in Figure~\ref{fig:pressure}, the pre-cluster cloud pressures
from both the {\Gizmo} and {\Gadget} simulations fall in the range of $P/k
\sim 10^8 - 10^{12}\, \rm{K cm^{-3}}$, in good agreement with
observations of a proto-SSC cloud in Antennae \citep{Johnson2015}, but they are $10^4 - 10^8$ times higher than
the typical pressure in the ISM \citep{Jenkins1983, Boulares1990, Welty2016}. 
Our results support the theoretical expectations that
massive SCs form in high-pressure clouds.

Moreover, the pressure distributions have near lognormal-shape
profiles in all panels. Such a quasi lognormal-shape pressure
distribution may be the cause of the quasi lognormal-shape ICMF. If we
assume a certain cluster formation efficiency $\eta$ ($\eta$ that varies
with galactic environment, from 0.01 in quiescent galaxies to $> 0.4$
in interacting galaxies, as suggested by \citealt{Goddard2010,
  Kruijssen2015}), then the mass of a SC $\rm {M_{cluster}}$
may be related to that of the birth clouds $\rm {M_{cloud}}$ as $\rm
{M_{cluster} \propto \eta M_{cloud}}$; then by inverting Equation~\ref{eqn:1}
we get $\rm {M_{cluster} \propto \eta M_{cloud} \propto \eta P}$. This
qualitative relation, as can be inferred from Figure~\ref{fig:pressure}, suggests that the 
distribution of cluster mass
depends on that of the cloud pressure, so a lognormal pressure
distribution may lead to a lognormal mass distribution of the
resulting clusters.

Furthermore, Figure~\ref{fig:pressure} also shows that the pressure
profiles have a peak at cluster mass around $10^{5.8 - 6}\, \Msun$ in the
{\Gizmo} simulation and around $10^{6 - 6.2}\, \Msun$ in the {\Gadget}
simulation, which are exactly the same peaks of their corresponding
mass functions.  This striking correlation may explain the preferred
mass range around the characteristic peak $10^{6}\, \Msun$, as determined by the cloud
pressure, for the newly formed SCs in galaxy mergers.

\begin{figure}
 \includegraphics[width=0.45\textwidth]{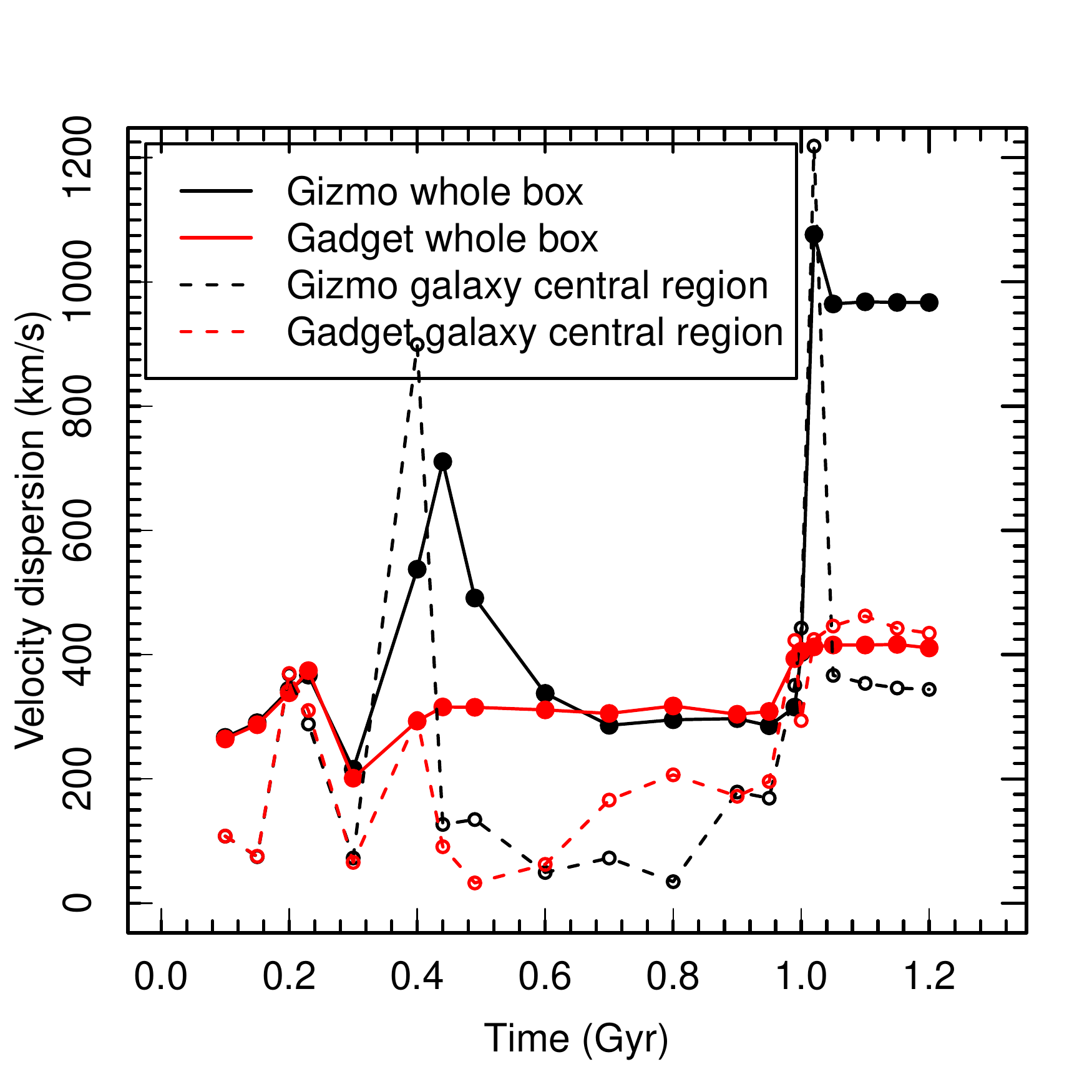}
 \caption{ Evolution of the total velocity dispersion of gas and stars in the central region of one galaxy ($5\times 5\times 5 \,\rm{kpc}^3$)
 (dashed) and the entire simulation box (solid) from both Gizmo (black) and Gadget (red) simulations, respectively.}
 \label{fig:shock}
\end{figure}

In order to understand the possible physical processes behind the high pressure of the pre-cluster clouds in our simulations, we compare the evolution of the total velocity dispersion of gas and stars in the central region of one galaxy (since the two merging galaxies are identical),  i. e., within 5 kpc from the galactic center, and that of the entire simulation box, as shown in Figure~\ref{fig:shock}. We find that the peaks of the total velocity dispersion correspond to those of the star formation as  in Figure~\ref{fig:SFR}, and that velocity dispersion of the galactic central region is higher than that of the whole box during the major starburst phases at 0.4 Gyr and 1 Gyr. These results suggest that high velocity dispersion around galaxy center, which leads to high pressure,  and strong circumnuclear starburst, may result from the same mechanism, compressive shocks driven by gravitational torques during galaxy merger.

Our theoretical findings may provide explanation to a number of observations. In addition to the recent observations of high pressure in a proto-SSC cloud in Antennae \citep{Johnson2015}, measurements of molecular clouds in the Antennae galaxies have revealed very high velocity dispersion in high star-forming regions \citep{Zhang2010}, which can be explained
by compressive shocks \citep{Wei2012}. \cite{Herrera2011} carried out near infrared imaging spectroscopy of the same region and found extended line widths in $\rm{H_2}$ emission, which indicates powerful shocks in the region. Similarly, measurements of the CO emission from the starbursting merger of M81 /M82 by \cite{Keto2005}  suggest that the molecular
clouds undergoing star formation are driven by shock compression. Theoretical studies \citep{Jog1992,  Ashman2001} have also shown that 
during the galaxy encounters, the giant molecular clouds undergo significant shock compression which leads to an increase in the cloud pressure. 

We also note that during $1a$ phase at 0.23 Gyr, the velocity dispersion of the central region is similar that of the whole box, which suggests a spatially extended star formation probably influenced by tidal forces, as discussed in $\S 3.1$. Simulation of Antennae galaxies by \cite{Renaud2014, Renaud2015} have shown that compressive tides during the galactic encounter can cause high star formation over extended volumes.

We can see from Figure~\ref{fig:pressure} that the pressures of our
simulated clouds are somewhat higher than that of the observed systems. This
can arise from the fact that in our simulation the two equal-mass,
Milky Way - size galaxies collide mainly head on and merge violently,
whereas the Antennae Galaxies have a smaller mass and they are on a
milder pericentric passage \citep{Renaud2015}. The extreme conditions
in the simulated galaxies produce more powerful shocks which in turn
help increase the gas pressure. Such an extreme high-pressure
environment may preferentially form massive SCs in a narrow
mass range as shown in Figure~\ref{fig:icmf}, which may help to explain
why we do not see a power-law ICMF in the simulations.

Our simulations bridge the observations and theories of cluster
formation and we confirm that massive SCs can form in gas-rich
galaxy mergers due to the high gas pressure produced by strong
gravitational interactions. Moreover, the special pressure range in
such merger environments preferentially forms SCs in a
narrow mass interval around the peak mass at  $\sim 10^{6}\, \Msun$. Furthermore, the
quasi-lognormal pressure distribution may lead to the quasi-lognormal
ICMF of SCs formed in colliding galaxies. Our results,
therefore, provide clues to the formation of globular clusters and
their universal lognormal mass functions, which will be explored in
the next section.

\section{Evolution of Massive Star Clusters}
\label{sec:evolution}

\begin{figure*}
\centering
\includegraphics[width=0.49\textwidth]{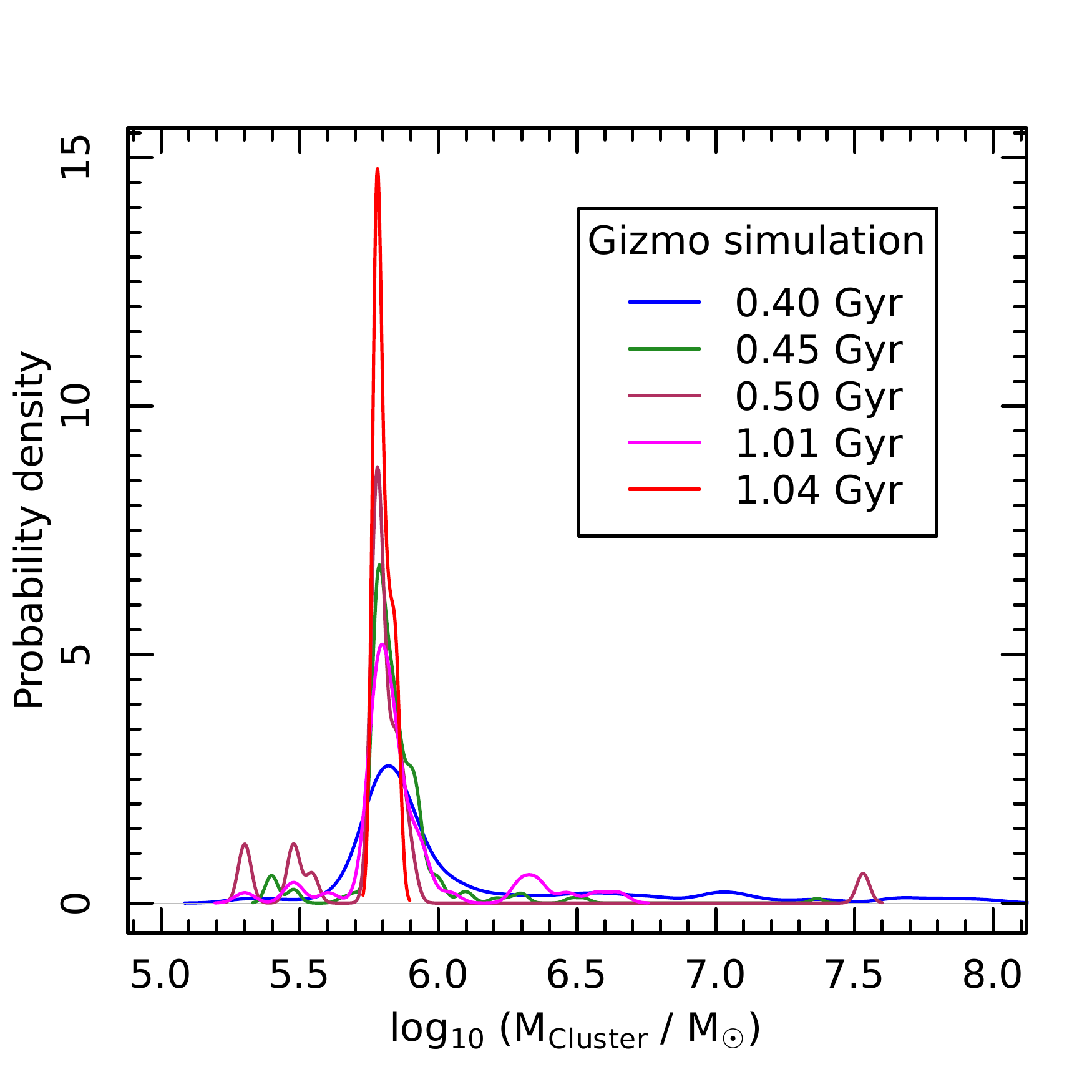}
\includegraphics[width=0.49\textwidth]{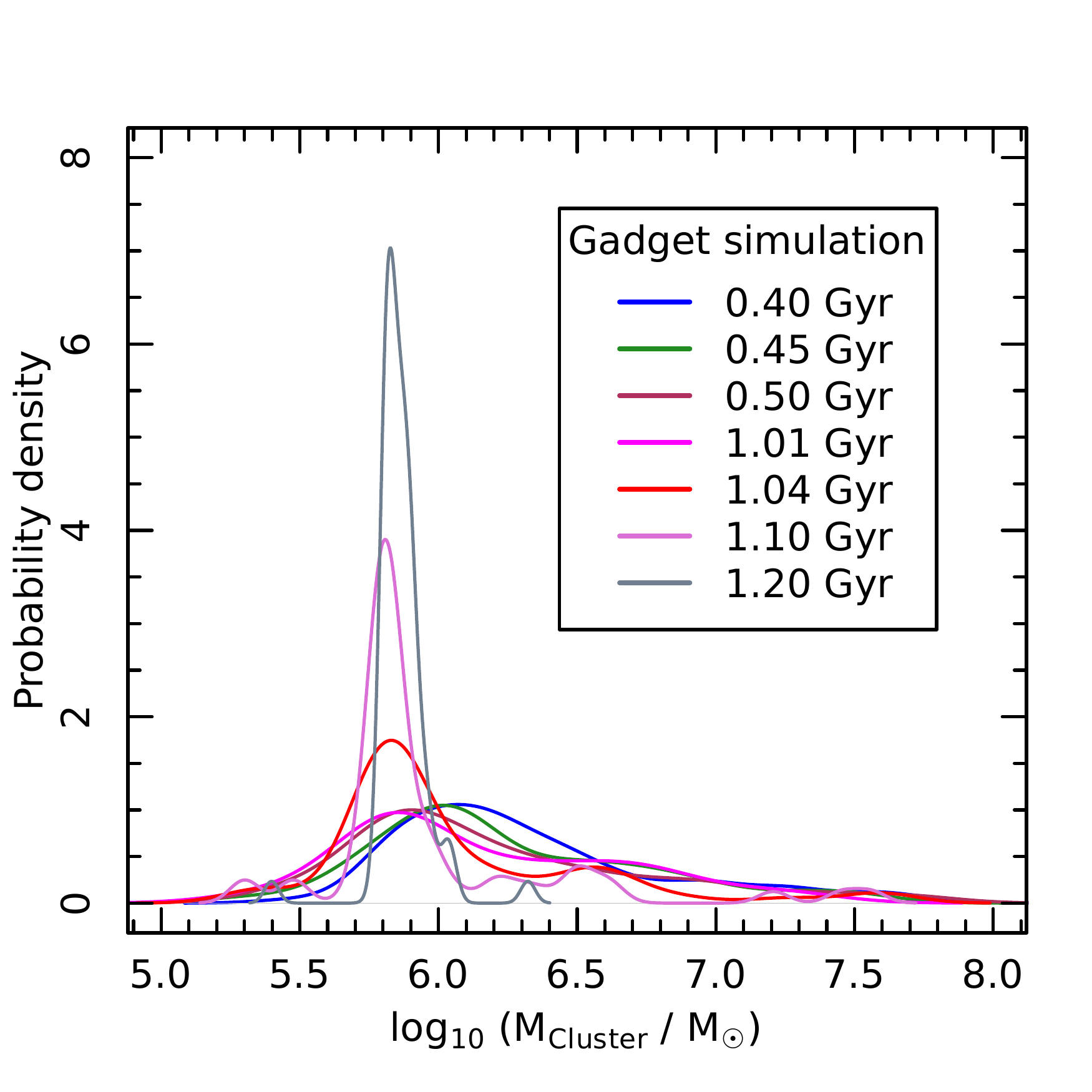}   
\caption{Probability density function of star cluster mass distribution from both {\Gizmo} (left panel) and {\Gadget} (right panel) simulations at 
different times, as indicated by the different colors. A Gaussian profile is used as the kernel density estimator. The PDFs are normalized such 
that the area under each curve is unity.} 
\label{fig:density}
 \end{figure*}

 \begin{figure}
 \includegraphics[width=0.49\textwidth]{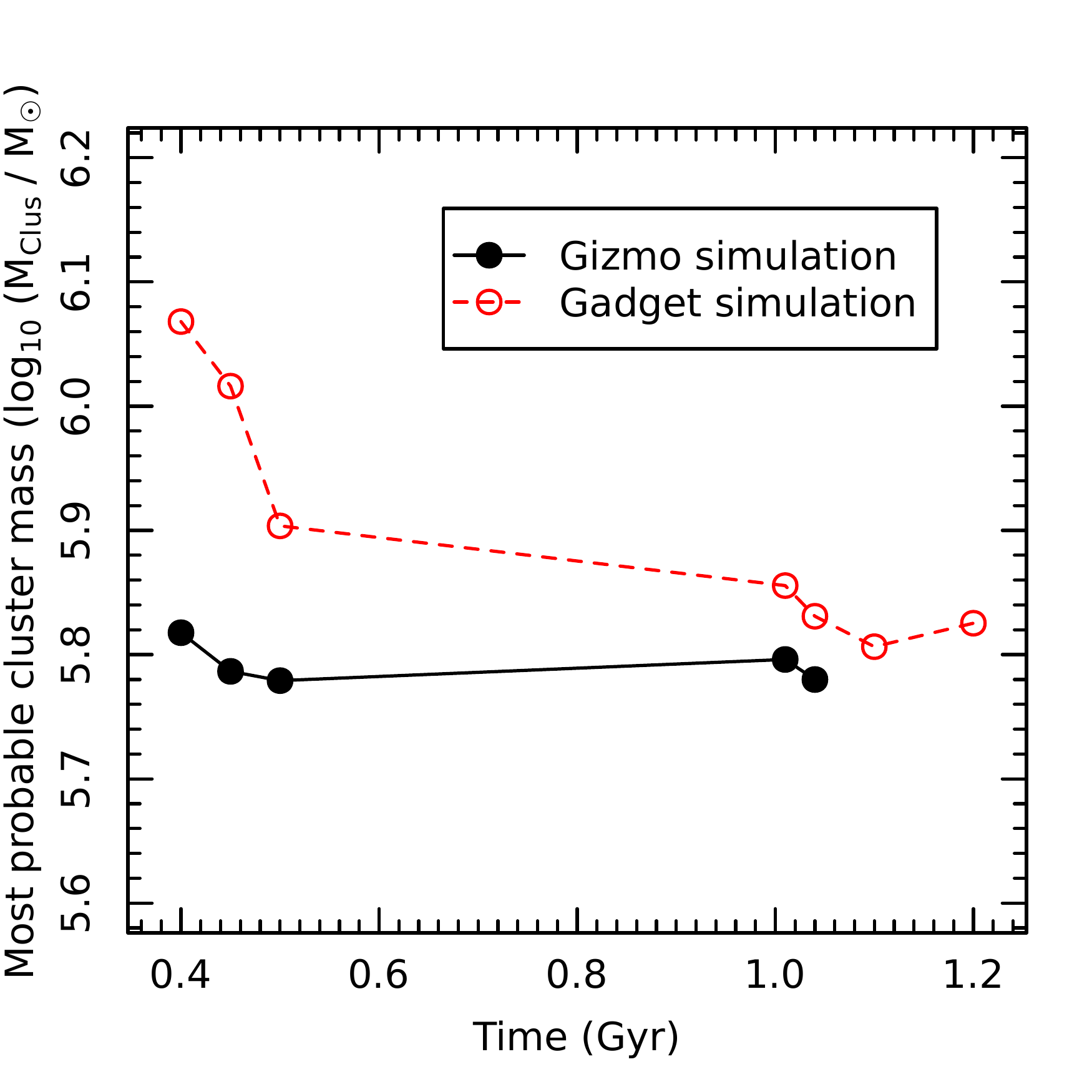}
\caption{The most probable star cluster mass as a function of time from both {\Gizmo} (black) and {\Gadget} (red) simulations. 
The data points correspond to the snapshot times in Figure~\ref{fig:density}.} 
 \label{fig:mpm}
\end{figure}

\begin{figure*}
 \centering
  \includegraphics[width=0.4\textwidth]{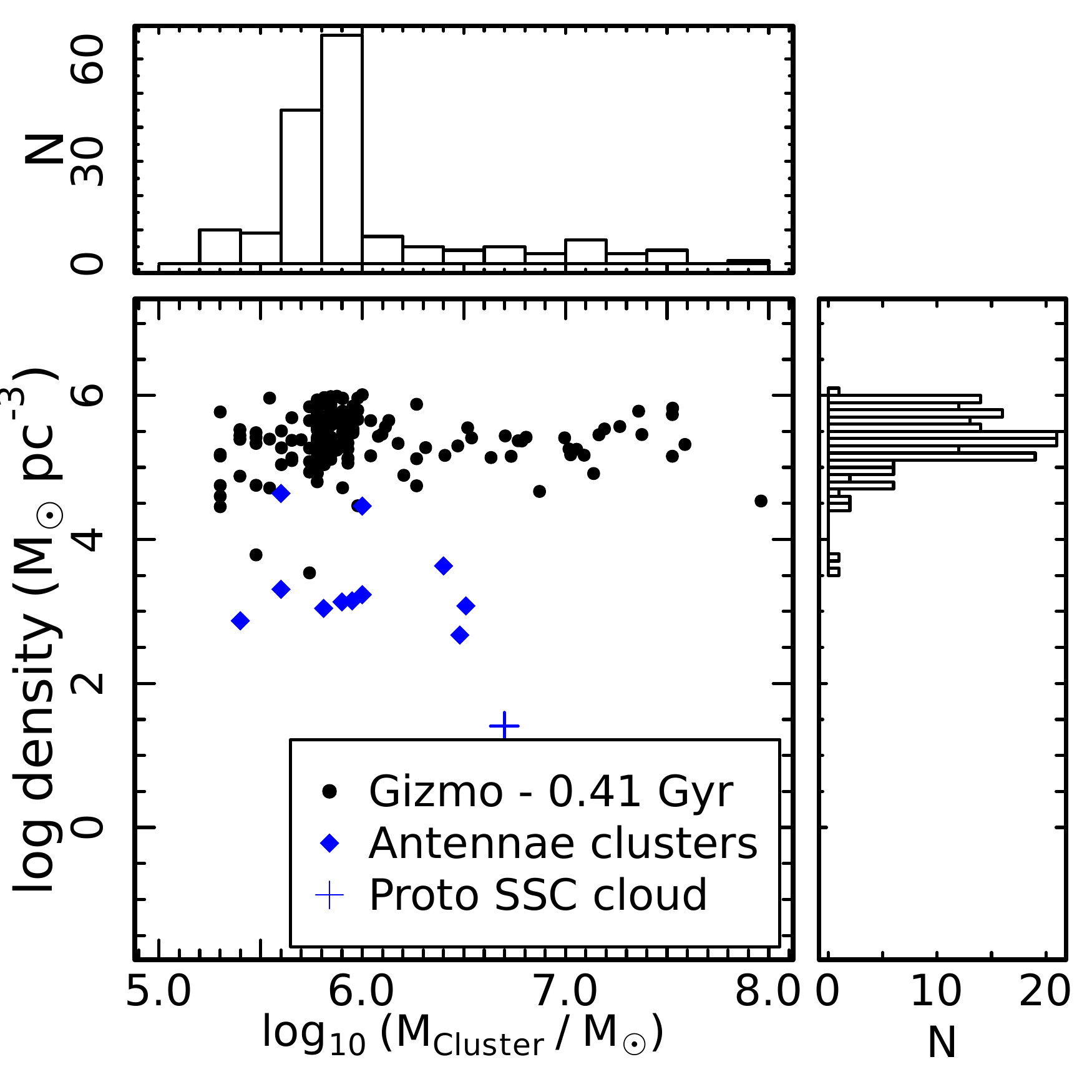}
   \includegraphics[width=0.4\textwidth]{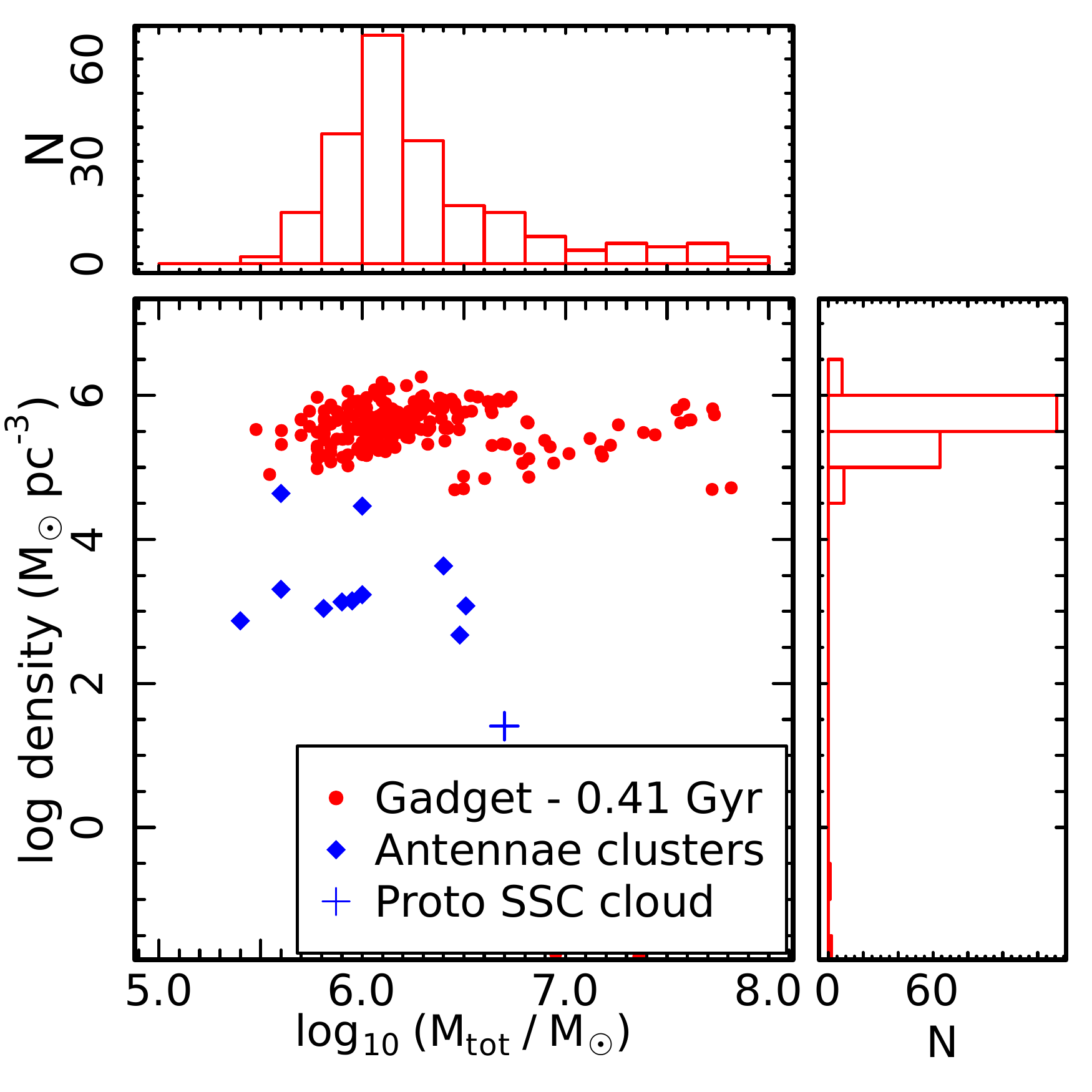}
   \caption{Density of the simulated clusters from both {\Gizmo} (black) and {\Gadget} simulations. The symbols follow the same 
   meaning as described in Figure~\ref{fig:pressure}.}
   \label{fig:density_clusters}
\end{figure*}

In order to track the change of the cluster mass functions over time,
we follow the evolution of the massive SCs in our
simulations up to 1.3 Gyr when the progenitors completely
coalesce. After the initial bursts of star and SC formation
during the first close passage at $t \sim 0.2 - 0.5$ Gyr, the activity
decreases, and many of the clusters are destroyed over time. However,
the starburst activity is renewed again at $t \sim 0.9 - 1.1$ Gyr
during the final coalescence, although at a lower amplitude than the
first burst.

Figure~\ref{fig:density} shows the probability density function (PDF)
of the cluster mass distribution at different times from both {\Gizmo}
and {\Gadget} simulations. Interestingly, the PDFs from the {\Gizmo}
simulation have similar narrow profiles, and the position of the peak
remains nearly the same over 1 Gyr, while those from the {\Gadget}
simulation show a slow evolution from a broad profile to a narrow one
and a shift of the peak mass by 0.3 dex over 1 Gyr. The narrowing of the
PDF is due to destruction of SCs at both low- and high-mass end by a variety of processes.
 After $\sim 1$ Gyr of evolution,
about one fourth of the SCs are left, most of them just around the peak mass. 
The difference in the narrowness of the PDFs from the two simulations is also
present in their initial mass functions in Figure~\ref{fig:icmf}. 
This probably stems from the different pressure distributions (Figure~\ref{fig:pressure}) owing to 
 different treatments of shocks between {\Gizmo} and {\Gadget}, as discussed in \S~\ref{subsec:codes}. 

The peak for each density curve is the value of the most probable mass
of the clusters for that time, as shown in Figure~\ref{fig:mpm}.  The
most probable cluster mass in the {\Gizmo} simulation is remarkably
consistent at $\sim 10^{5.8}\, \Msun$ over 1 Gyr, while that of the
{\Gadget} simulation changes slightly from $\sim 10^{6.09}\, \Msun$ at
the initial starburst to $\sim 10^{5.8}\, \Msun$ after 1 Gyr. Since
these are isolated merger simulations, it is meaningless to continue
the simulations for a longer time, but in a cosmological context, it
can be predicted that the SCs will lose some of their mass
due to stellar evolution and other destructive
processes. Semi-analytical and N-body simulations by
\cite{Kruijssen2015} and \cite{Webb2015} found that clusters can lose
mass by a factor of $2-4$ after a Hubble time of evolution.  The
eventual mass of our simulated clusters can be similar to the
observed peak mass of globular clusters at $1.5 - 3 \times 10^5\,
\Msun$ \citep{Harris2001, Jordan2007}.

These results show that the shape of the mass function and the position of the mass peak of massive clusters have little evolution 
over the course of the galaxy collision of more than 1 Gyr.  We note that the evolution of star clusters is subjected to several 
destructive processes \citep[e.g.,][]{Gnedin1999a,Fall2001}. For
low-mass clusters ($< 10^5\,\Msun$), the destruction is mainly dominated by two-body relaxation processes, in which the mass of a cluster   
linearly decreases with time until it is destroyed. For more massive clusters, the evolution is primarily influenced by stellar evolution
at early times ($\lesssim 100$ Myr) and by gravitational shocks at later times. These effects are included in our hydrodynamic simulations
but the mass resolution is not high enough to resolve the processes realistically, since the two-body relaxation and stellar evolution 
depend on individual stars. 
However, the cluster disruption time-scale due to two-body relaxation is proportional to the cluster mass, 
$t_{rlx}\sim 1.7\,\rm{Gyr}\,\times(\rm{M_{clus}}/10^4\Msun)^{0.62}\times(T/10^4\,\rm{Gyr}^{-2})^{-0.5}$, 
where $M_{clus}$ is the cluster mass and $T$ is the tidal strength around the clusters \citep{Kruijssen2012b}.
Using the minimum cluster mass of $2\times10^5\,\Msun$ in our simulation, and a typical range of tidal strength
in the nuclear region (since most of these clusters are concentrated around galaxy nuclei) 
$T \sim 0.1 - 50 \times10^{-30}\,s^{-2}$ \citep{Renaud2010}, we find that the range of the disruption 
time-scale is $t_{rlx}\sim 4.88 - 108.8$ Gyrs. For more massive SCs, the time-scale is even longer,  beyond
our run time of 1 Gyr. Therefore, the two-body relaxation may not have a major disruptive effect on these SCs.
Furthermore, the mass loss time-scale due to gravitational shocks
($t_{sh}$) depends strongly on the cluster density, $t_{sh} \sim 3.1\,\rm{Gyr}\times\rho/10^4\, \Msun/pc^3$ \citep{Kruijssen2012b}.
The majority of our clusters 
have  a density range of $\rho \sim 10^{5} - 10^{6}\,  \Msun/pc^3$, as shown in Figure~\ref{fig:density_clusters}, which suggests 
$t_{sh} \sim 30 - 300$ Gyrs, much longer than the Hubble time. So gravitational shocks may not have a significant impact on the 
clusters in our simulation. In addition, as demonstrated by \cite{Renaud2013}, star clusters formed in galaxy mergers are also 
affected by the intense tidal field of the galaxies, more so for clusters in the merger remnant compared to the ejected ones. We
note, however, the clusters in their simulations have masses $\lesssim 3\times10^4\Msun$, and mass loss decreases as the cluster
mass increases. For example, for a cluster to increase its mass by a factor of 2, from $1.6\times 10^4\, \Msun$ to $3.2\times 10^4\, \Msun$,
its survival rate (fraction of initial mass survived) after 1 Gyr increases from 0.6 to 0.7. Extrapolating this trend to our clusters which
are $\sim 10$ times more massive than those in \cite{Renaud2013}, we argue that destruction from tidal fields likely has negligible effects
on the clusters we consider here.

Our results suggest that the observed globular clusters may form in high pressure environments induced by galaxy 
interaction at high redshift when the merger rate was high \citep[e.g.,][]{Rodriguez2015, Mistani2016}. The extremely
high gas pressures in the merging environments produce lognormal ICMFs with a peak mass around $\sim 10^{6}\, \Msun$,
and they evolve slowly over a Hubble time into the universal lognormal
profiles with a peak at $\sim 1.5-3 \times 10^{5}\, \Msun$ as observed
today.

\section{Discussion}
\label{sec:discussions}

We have used the two codes {\Gadget} and {\Gizmo} with significantly
different hydrodynamical solvers to simulate the formation and
evolution of SCs in galaxy mergers and found similar
results.  This helps us to reduce the possibility of significant numerical
artifacts in our results and suggests that our findings are physical
and robust.

In {\Gadget}, the hydrodynamic solver uses a smoothing scheme
\citep{Springel2005}, where the physical properties are averaged over
a given number of neighboring particles, which is 32 in our
simulation.  The mass of 32 star particles for our resolution is close
to the peak cluster mass seen in the {\Gadget} simulation. However,
the {\Gizmo} simulation has no smoothing procedure and the properties
do not depend on the number of neighbors but we still see the cluster
mass peak at similar masses. This suggests that the peak cluster mass
found in our simulation is likely not a numerical artifact, but rather
may be a physical feature of SCs formed in such extreme
environments.  One very potent source of the high pressure in our
simulations is the shocks produced during the close passages of
galaxies and starburst periods.  {\Gizmo} handles shocks much better
than {\Gadget}, and it calculates the effects of contact
discontinuities more precisely, captures fluid mixing instabilities
well and has less numerical noise. These differences are pronounced in
Figure \ref{fig:pressure}, where the pressure distribution is
more cleanly peaked in {\Gizmo} whereas it is quite spread out in
{\Gadget}.

In our simulations, feedback mechanisms from supernovae and active galactic nuclei in the form of thermal energy and galactic winds are included. 
However, other feedback processes such as photoionization and radiative pressure may affect star formation \citep{Krumholz2014}. 
On the one hand, high-energy UV photons from OB stars can ionize the HII clouds in ISM, the expanding  HII clouds can compress the neutral gas in 
the outskirts of molecular clouds and the fragmentation of these dense gas can increase the star formation (positive feedback). On the other hand,
the momentum imparted on HII gas by ionizing photons can drive gas out of the central regions of GMCs which may suppress star formation or unbind 
star clusters (negative feedback).  Simulations by \cite{Dale2005} show that for very dense clouds (core density $\sim 10^8\,\rm{cm}^{-3}$), a highly
collimated gas outflow can carry the extra momentum out of the cloud without unbinding the cluster. The photoionization also drives the Jeans mass
down, resulting in higher star formation. In the context of star formation in galaxy mergers, comparative studies of merger simulations with and
without these feedback processes by \cite{Hopkins2013} showed that detailed feedback promote more extensive star formation in tails and bridges,
but the global star formation properties remain similar. We plan to explore the effects of radiative feedback on the formation and properties of
star clusters in a future project.

We have seen from Figure \ref{fig:icmf} that the ICMF
in our simulations does not have the shape of a falling power
law, which is commonly observed for many YMC systems. Rather, the ICMF
has a quasi-lognormal shape which is preserved in later stages. We
doubt that this is due to the limited mass resolution ($5\times
10^4\,\Msun$) in our simulations, as similar ICMFs were also reported
by \cite{Renaud2015}, who performed a simulation of the Antennae using
a grid-based AMR code with very high mass resolution ($\sim 70\,
\Msun$).  The agreement among the merger simulations using different
codes, various feedback prescriptions and across resolution suggest
that the lognormal-shape ICMFs and the unique mass peak are mostly
likely special features of clusters formed in the extremely high
pressure environments produced by galaxy collisions. In fact, some
studies of active galaxies, such as the starburst galaxy NGC 5253
\citep{Cresci2005} and the interacting Antennae pair
\citep{Anders2007} have shown that the ICMF for YMCs is not a power
law for all mass scales, but may rather have a turnover at low mass.

As indicated by Figure \ref{fig:density} and Figure \ref{fig:mpm},
over the evolution of more than 1 Gyr, the mass functions of our star
clusters in both the {\Gadget} and {\Gizmo} simulations survives
destructive processes and retain the same quasi-lognormal shape with a
consistent peak at around $\sim 10^{5.8}\, \Msun$.  This mass is quite
close to the observed GCMF peak at $\sim 1.5 -3\times 10^{5}\, \Msun$. Ideally we
would like to have a fully cosmological hydrodynamic simulation of
galaxy formation and evolution with a very high mass resolution ($\sim
10^{3}\, \Msun$) to identify SCs and evolve them for $\sim 13$ Gyr
to explore the fate of the globular cluster mass function, but that
remains computationally very expensive.  However, the strong trend of
survival of the lognormal shape of the ICMFs in our simulations lends
support to our speculation that the origin of the lognormal mass
functions of the globular clusters may come from the extremely high
pressure formation conditions in interacting galaxies.

\section{Conclusions}
\label{sec:conclusions}

We have performed high-resolution hydrodynamic simulations of galaxy
mergers using two different codes and studied the formation and
evolution of SCs in them. We obtained consistent results
from both codes, suggesting that our results are physical and
robust. Here is a summary of our findings:
 
\begin{itemize} 

\item A strong galaxy interaction produces intense shocks and
compression of gas, which triggers global starbursts and the
formation of massive SCs in the nuclear regions of the
mergers and in the tidal bridges and tails. The massive star
clusters show quasi-lognormal ICMFs in the
range of $\sim 10^{5.5-7.5}\, \Msun$ with a peak around $10^6\, \Msun$.
 
\item The nascent cluster-forming gas clouds have very high pressures
in the range of $P/k \sim 10^{8-12}\, \rm{K\, cm^{-3}}$, in good
agreement with observations and theoretical expectations that
massive SCs form in high-pressure environments, which 
naturally arise in violent galaxy collisions. Moreover, the gas pressures
show quasi-lognormal profiles, which suggest that the
quasi-lognormal ICMFs of the clusters may be caused by the pressure
distributions in the birth clouds. Furthermore, the peak of the
pressure distribution correlates with the peak of the cluster mass
function at $10^{5.8-6.2}\, \Msun$, indicating that clusters
formed in such extremely high pressure clouds have a characteristic mass
around $\sim 10^{6}\, \Msun$.
 
\item The cluster mass functions evolve slowly over time with a
  declining cluster number due to destructive processes, but the quasi-lognormal shape and the peak of the mass 
  functions do not change significantly during the course of galaxy collisions over 1 Gyr. 
 
\end{itemize}

Our results suggest that the observed universal lognormal globular
cluster mass functions and the unique peak at $\sim 2\times 10^{5}\,
\Msun$ may originate from the high-pressure formation conditions in
the birth clouds. Globular clusters may have formed in extremely high
pressure environments produced by violent galaxy interactions at high
redshift when mergers were more common. The lognormal cluster mass
functions with a preferred most probable cluster mass around $\sim
10^{6}\, \Msun$ may be unique products of such extreme birth
conditions, and they evolve slowly over 13 Gyrs but retain the
lognormal shape and peak against destructive processes.

\section*{Acknowledgement}
We thank Dr. Bruce Elmegreen and Dr. Phil Hopkins for valuable discussions, and Dr. Florent Renaud for 
his constructive comments which have helped improve the paper significantly. YL acknowledges support 
from NSF grants AST-0965694, AST-1009867, AST-1412719, and MRI-1626251.  
AK is supported by the {\it Ministerio de Econom\'ia y Competitividad} and the
{\it Fondo Europeo de Desarrollo Regional} (MINECO/FEDER, UE) in Spain through
grant AYA2015-63810-P as well as the Consolider-Ingenio 2010 Programme of the
{\it Spanish Ministerio de Ciencia e Innovaci\'on} (MICINN) under grant
MultiDark CSD2009-00064. He also acknowledges support from the {\it Australian
Research Council} (ARC) grant DP140100198.
We acknowledge the NSF award MRI-1626251 for providing computational resources and services through Institute for CyberScience at The Pennsylvania State
University that have contributed to the research results reported in this paper. The
Institute for Gravitation and the Cosmos is supported by the Eberly
College of Science and the Office of the Senior Vice President for
Research at the Pennsylvania State University.

\bibliography{sc_v6}
\end{document}